\documentclass[a4paper,11pt]{article}
\pdfoutput=1
\usepackage{jheppub} 

\usepackage[T1]{fontenc} 
\usepackage{graphicx,xcolor}
\usepackage{float}
\usepackage{bbm}
\usepackage{slashed}
\usepackage{mathtools,braket,blkarray}
\allowdisplaybreaks
\usepackage{amssymb,amsfonts,relsize}
\usepackage[width=0.9\textwidth,footnotesize]{caption}
\usepackage[font=scriptsize]{subcaption}
\hypersetup{pageanchor=false}
\usepackage{booktabs,multirow,makecell}
\usepackage[utf8]{inputenc}
\usepackage{cleveref}

\newcommand{\overbar}[1]{\mkern 1.5mu\overline{\mkern-1.5mu#1\mkern-1.5mu}\mkern 1.5mu}

\newcommand*\diff{\mathop{}\!\mathrm{d}}
\preprint{IFT-UAM/CSIC-20-145}
\arxivnumber{2011.08195}

\title{Dark Matter candidates in a Type-II radiative neutrino mass model}
\author[a]{Roberto~A.~Lineros}
\author[b,c]{and Mathias~Pierre}

\affiliation[a]{Departamento de Física, Universidad Católica del Norte, Avenida Angamos 0610, Casilla 1280, Antofagasta, Chile.}
\affiliation[b]{Instituto de Física Teórica (IFT) UAM-CSIC, Campus de Cantoblanco, 28049 Madrid, Spain.}
\affiliation[c]{Departamento de Física Teórica, Universidad Autónoma de Madrid (UAM), Campus de Cantoblanco, 28049 Madrid, Spain.}

\emailAdd{roberto.lineros@ucn.cl}
\emailAdd{mathias.pierre@uam.es}

\abstract{
We explore the connection between Dark Matter and neutrinos in a model inspired by radiative Type-II seessaw and scotogenic scenarios. In our model, we introduce new electroweakly charged states (scalars and a vector-like fermion) and impose a discrete $\mathbb{Z}_2$ symmetry. Neutrino masses are generated at the loop level and the lightest $\mathbb{Z}_2$-odd neutral particle is stable and it can play the role of a Dark Matter candidate. We perform a numerical analysis of the model showing that neutrino masses and flavour structure can be reproduced in addition to the correct dark matter density, with viable DM masses from 700 GeV to 30 TeV. We explore direct and indirect detection signatures and show interesting detection prospects by CTA, Darwin and KM3Net and highlight the complementarity between these observables. }

\begin{document}
\maketitle
\flushbottom
\section{Introduction}
\label{sec:intro}

The presence of Dark Matter (DM) and its role in the formation of large scale structures of the universe~\cite{Aghanim:2018eyx} altogether with the observation of neutrino oscillations~\cite{Whitehead:2016xud, Decowski:2016axc, Abe:2017uxa, Capozzi:2016rtj, Esteban:2018azc, deSalas:2017kay} are some of the strongest indications that the Standard Model (SM) of particle physics lacks essential ingredients. Indeed, the SM provides a successful description of the microscopic interactions with a remarkable accuracy. However, neutrinos, which are precisely massless within the SM, should possess non-vanishing masses in order to explain successfully the observed oscillation patterns. In addition, the SM do not include a viable dark matter candidate among its large particle content. Both topics have became the ground base motivation for a large variety of the Beyond-the-Standard-Model (BSM) constructions.  

As stated before, neutrino oscillations require massive neutrinos and therefore the SM must be extended in order to accommodate non-vanishing masses. One of most famous and minimal mechanisms accounting for non-vanishing neutrino masses is known as the ``see-saw" mechanism~\cite{Minkowski:1977sc, Yanagida:1979as, Mohapatra:1979ia, Schechter:1980gr, Schechter:1981cv, Foot:1988aq}. In seesaw-based models, extra fields such as $SU(2)_L$ fermion singlets, triplets, scalar triplet, with or without hypercharge, are added to the particle content of the SM, allowing to generate neutrino Majorana mass terms. On the effective point of view, the lowest dimensional operator responsible for Majorana mass terms is the Weinberg operator which is a dimension-5 operator constructed with SM fields that preserves every SM gauge symmetries but violate lepton number conservation by 2 units: 
\begin{equation}
    \mathcal{O}_5 \propto \dfrac{1}{\Lambda}(L_i H)^T (L_j H) \, ,
\end{equation}
where $L_i$ are the lepton doublets, $H$ is the SM higgs doublet, and $\Lambda$ is a mass scale. The Majorana mass term arises after the $SU(2)_L \times U(1)_Y$ spontaneous symmetry breaking (SSB):
\begin{equation}
m_\nu \sim v^2/\Lambda\,.
\end{equation}
Within the BSM models landscape, the Weinberg operator is typically generated by integrating out of the spectrum heavy mediators. Models reproducing light neutrino Majorana masses at tree-level by adding a minimal number of extra fields can be generally sorted in three main categories. These categories are commonly referred to as Type-I, Type-II, and Type-III seesaws (see~\cite{Lattanzi:2014mia} for an overview). Among them, diagrams responsible for the Weinberg operator in T-I and T-III seesaw models are similar. Within the T-I seesaw framework, a right-handed neutrino (Majorana fermion $SU(2)_L$ singlet) is introduced instead of the Majorana fermion $SU(2)_L$ triplet used in the T-III. T-II seesaw is different as a $SU(2)_L$ scalar triplet with hypercharge is added in this case. This field carries 2 units of lepton number, which is broken spontaneously once this scalar acquires a non-vanishing vacuum-expectation-value (vev), generating subsequently neutrino mass terms proportional to this vev.

Beyond this framework, non-vanishing neutrino masses can also be generated at the loop level, providing an additional argument to justify the large hierarchy between the electroweak scale and the light neutrino masses. For these cases, an interplay between neutrino masses and DM presents interesting features. First of all, tree-level processes could become negligible or forbidden by invoking an additional discrete or gauge symmetry~\cite{Ballett:2019cqp,Gehrlein:2019iwl}. For instance the simplest realization is typically obtained by considering a $\mathbb{Z}_2$ discrete symmetry. This symmetry also protects one-loop processes, making them the only contributors to the neutrino masses. As a side effect of this symmetry, particles running inside the loop are protected, among these, the lightest state is stable and could be a viable DM candidate. There are several scenarios considering DM candidates in this framework (see for instance~\cite{Restrepo:2013aga}). However, we highlight the models known as ``scotogenic'' , in which, the Weinberg operator is generated at one-loop by using as basis the Type-I~\cite{Ma:2006km} or Type-III~\cite{Ma:2008cu} seesaw BSM fields which are, in this case, charged under the new symmetry. Constructions combining both models present interesting DM and neutrino phenomenology (for instance~\cite{Suematsu:2019kst, Hirsch:2013ola,Avila2020,Avila:2019hhv}). Nevertheless, models using the BSM fields of the Type-II seesaw are less common due to the ambiguity in their definition~\cite{Kanemura:2012rj,Lu:2016dbc,Chen:2019okl}. In addition, some of the phenomenological aspects are not explored.
In this paper we investigate the phenomenology of a model accounting both for the observed dark matter abundance in our universe in addition to light neutrino masses, based on a Type-II seesaw mechanism.

Our manuscript is constructed as follows: in Section~\ref{sec:model}, we provide a description of various aspects of the model, such as the particle content and corresponding charge assignments. In Section~\ref{sec:neutrinodarkmatter}, we explore the neutrino physics, DM phenomenology and lepton flavour violating observables implied by the model. The results of an exhaustive scan of the parameter space is presented in Section~\ref{sec:numerical_analysis} which is subsequently analyzed and commented. We consider further constraints and discuss additional relevant observables in Section~\ref{sec:addconstraints} before concluding in Section~\ref{sec:conclusions}. \par \medskip

\section{The Model}
\label{sec:model}
In addition to the SM particle content, we introduce four new fields charged under a discrete $\mathbb{Z}_2$ symmetry with non-trivial charge assignment under the SM gauge symmetry group: a hyperchargeless $Y=0$ real $SU(2)_L$ triplet scalar $\Omega$, a hypercharged $Y=1$ complex $SU(2)_L$ triplet scalar $\Delta$ and two fermionic $SU(2)_L$  doublets $f_{L,R}$ with opposites chiralities but identical quantum numbers. The SM particle content is uncharged under the discrete $\mathbb{Z}_2$ symmetry or equivalently $\mathbb{Z}_2$-even. The field content of this model is summarized in Table~\ref{tab:charges}. The $\mathbb{Z}_2$ discrete symmetry 
will play a two-folded role here.
On the first hand, as in the original scotogenic model~\cite{Ma:2006km}, neutrino masses are generated at the loop-level, by receiving contributions of fields charged under a discrete $\mathbb{Z}_2$ symmetry. On the other hand, the neutral fields charged under the $\mathbb{Z}_2$ will render a suitable stable Weakly-Interacting-Massive-Particle (WIMP) DM candidate. There are two meaningful differences arising from the original proposal~\cite{Ma:2006km}. The inclusion of one triplet field with $Y=1$ and also charged under a $\mathbb{Z}_2$ symmetry
comes with doubly charged scalars which must decay in pairs of particles that are also charged under the $\mathbb{Z}_2$,
thus forbidding the well-known smoking gun prospect of $\Delta^{++}\longrightarrow \ell_\alpha^+ \ell_\beta^+$
existing in models extending the TII-seesaw~\cite{Farzan:2010mr}. The inclusion of a second triplet also charged under the $\mathbb{Z}_2$, but with $Y=0$ renders two features. Firstly, it gives rise to a loop made-up of charged particles contributing to neutrino masses. This comes in a similar way as it comes in Bilinear R-parity violation, where charged loops are fundamental in order to give rise to a full description of neutrino masses.
Secondly, this second triplet also renders a richer scalar $\mathbb{Z}_2$-odd sector, with different features as other scotogenic models, for instance the model of Ref.~\cite{Hirsch:2013ola}.
\begin{table}[b!]
    \centering
    \begin{tabular}{ |>{\centering\arraybackslash  \bf}m{0.16\textwidth} |*{6}{ >{\centering\arraybackslash}m{0.08\textwidth} |}}
        \hline Field & $L_i$ & $f_L$  & $f_R$ & $\Delta$ & $\Omega$ & $H$ \\
    	\hline 	\hline Spin  & 1/2 & 1/2 & 1/2 & 0 & 0 & 0   \\
    	\hline Chirality & \text{L} & \text{L} & \text{R} & -- & -- & --   \\
        \hline SU(2)$_{L}$ & $\mathbf{2}$ & $\mathbf{2}$ & $\mathbf{2}$ & $\mathbf{3}$ & $\mathbf{3}$ & $\mathbf{2}$  \\
        \hline U(1)$_{Y}$ & -1/2 & 1/2 & 1/2 & 1 & 0 & 1/2 \\
        \hline $\mathbb{Z}_2$ & +1 & -1 & -1 & -1 & -1 & +1  \\
        \hline
    \end{tabular}
    \caption{Charge assignment of the new  fields considered in this model in addition to the SM $i-$th flavour lepton doublet $L_i$ and SM Higgs doublet $H$.}
    \label{tab:charges}
\end{table}
The Lagrangian allowed by gauge invariance can be parametrized as
\begin{equation}
    \mathcal{L} \supset -y^i_\Delta \Big( \overline{f_R}  \Delta L_i +\text{h.c.} \Big)
    -y_\Omega^i \Big( \overline{f_L^c} i \sigma_2 \Omega L_i +\text{h.c.} \Big) -m_{f} \Big( \overline{f_L} f_R + \overline{f_R} f_L \Big) - V_\text{scalar}\,,
\end{equation}
where $L_i$ with $i=1,2,3$ denote the $i-$th flavour SM lepton $SU(2)_L$ doublet. $y^i_{\Delta,\Omega}$ are dimensionless free parameters carrying a flavour index $i$ and $m_f$ is a free mass parameter. In this work we assume that the new terms are CP-conserving and therefore we consider the Yukawa couplings as real parameters. $^c$ denotes the Lorentz charge conjugation operation and $V_{\rm scalar}$ is the scalar potential is explicited in Sec.~\ref{sec:scalar_sector}. Notice that the absence of right-handed neutrinos implies the absence of couplings between lepton doublets $L_i$ and the SM Higgs. On the other hand, the new triplets are coupled, via Yukawa terms, to the neutrinos. Their components can be parametrized as
\begin{equation}
  \Omega = \sum_{i=1}^3 \Omega_i \sigma_i =  \left( \begin{array}{cc}
    \Omega^0 &  \sqrt{2}\Omega^+ \\
    \sqrt{2} \Omega^- & -\Omega^0
  \end{array} \right)\,,\qquad \text{and} \qquad
  \Delta =  \sum_{i=1}^3 \Delta_i \sigma_i =  \left( \begin{array}{cc}
    \Delta^{+} & \sqrt{2}\Delta^{++} \\
     \Delta^{0} & -\Delta^{+}
  \end{array} \right)\,.
\end{equation}
where the $+,-$ exponents on the fields represent their electric charges. As discussed further on, such $\mathbb{Z}_2$-odd charged and neutral scalar fields induce radiative contributions to neutrino masses. In the following we provide a more detailed description of the fermionic and scalar sector of the model.

\subsection{Fermion sector}
\label{sec:fermionsector}
The new fermionic $SU(2)_L$ doublets can be parametrized as $f_{L,R}^T=(f_{L,R}^+,f^0_{L,R})$. At tree-level since only Dirac mass terms are present in the new fermionic sector, the two doublets with opposite chiralities combine to form Dirac particles $f=f_L+f_R=(f^+,f^0)^T$ with $f^+=f_L^+ +f_R^+$ and  $f^0=f_L^0 +f_R^0$ which are the mass eigenstates with respective charges $Q=1$ and $Q=0$. In term of these fermionic mass eigenstates the Lagrangian at tree-level can simply be written as
\begin{equation}
    \mathcal{L} \supset -y^i_\Delta \Big( \overbar{f}  \Delta L_i +\text{h.c.} \Big)
    -y_\Omega^i \Big( \overbar{f^c}  i \sigma_2  \Omega L_i +\text{h.c.} \Big) -m_{f} \Big( \overbar{f^+} f^+ + \overbar{f^0} f^0 \Big) - V_\text{scalar}\,.
    \label{eq:Lag_fermion}
\end{equation}
Notice here that the $Q=1$ and $Q=0$ fermionic states are degenerate at tree-level with masses $m_{f^+}=m_{f^0}=m_f$. However, loop corrections give rise to a mass splitting $\Delta m_f = m_{f^+} - m_{f^0}$ between these two components such that the mass terms in Eq.~(\ref{eq:Lag_fermion}) can be written at the one-loop level as
\begin{equation}
    \mathcal{L} \supset -m_{f} \left(  \overbar{f^0} f^0 + \Big(1+\dfrac{\Delta m_f}{m_ f}\Big)  \overbar{f^+} f^+  \right) \, .
\end{equation}
The relative mass splitting is typically small $\Delta m_f/m_f \lesssim 1\%$ for the relevant part of the parameter space, as shown further on. For instance $\Delta m_f\sim 10^{-4} \,m_f$ for $m_f=1$ TeV. The analytical expression for the mass splitting at the one-loop level $\Delta m_f$ can be found in Appendix~\ref{sec:fermion_mass_splitting}. 

\subsection{Scalar Sector}
\label{sec:scalar_sector}
The scalar sector of this model is composed of the Higgs field $H$, a $Y=1$ triplet $\Delta$ and a $Y=0$ triplet $\Omega$. The only $\mathbb{Z}_2$-even scalar is the Higgs doublet $H$ and it is responsible of the electroweak spontaneous symmetry breaking. In the following, we explicit all the terms present in the scalar-sector Lagrangian as well as the various mass eigenstates and eigenvalues, obtained after diagonalization of the scalar mass matrix.

\subsubsection{Kinetic terms}
The kinetic terms for the scalar triplets can be written as
\begin{equation}
    \mathcal{L}_\text{kin}=\dfrac{1}{2}\text{Tr}\left[ (D_\mu \Delta)^\dagger (D^\mu \Delta) \right]+\dfrac{1}{4}\text{Tr}\left[ (D_\mu \Omega)^\dagger (D^\mu \Omega) \right]\,,
\end{equation}
where the covariant derivatives are defined as
\begin{equation}
    D_\mu \Delta = \partial_\mu \Delta - i g \left[ \dfrac{\sigma_a}{2} W_\mu^a ,\Delta \right]-g^\prime B_\mu \Delta\,,
\end{equation}
and
\begin{equation}
    D_\mu \Omega = \partial_\mu \Omega - i g \left[ \dfrac{\sigma_a}{2} W_\mu^a ,\Omega \right]\,.
\end{equation}
using standard notations from the literature. We parametrize the neutral scalar fields as a function of real scalars
\begin{equation}
    \Delta^0= \Delta^0_R + i\Delta^0_I ~,~~    \Omega^0=\Omega^0_R~,
\end{equation}
where $\Delta^0_R, \Omega^0_R$ and $\Delta^0_I$ denote the real and imaginary parts of $\Delta^0$ and $\Omega^0$.

\subsubsection{Scalar Potential}
In this section we explore the scalar potential and the mass eigenstates arising from this sector.  First, a comment regarding the Higgs field. Since it is the only $SU(2)$ and $\mathbb{Z}_2$-even field, it constitutes the only source of electroweak symmetry breaking, and thus the tadpole equations will not be trivially satisfied which implies a non-zero vev. The resulting neutral pseudoscalar mode obtained after expanding around the minimum will be the Goldstone mode responsible of the longitudinal modes of the $Z$-boson. In the same footing, the charged component of the Higgs field will constitute the longitudinal mode of the $W$-bosons. Thus, all the physical scalar fields within this model but the Higgs field are $\mathbb{Z}_2$-odd. The full scalar potential of the theory, including all renormalizable terms allowed by gauge invariance, can be parametrized as
\begin{align}
  \label{eq:scalarpotential}
  V_\text{scalar} \, = \, & -\mu_h^2 |H|^2+\lambda_h |H|^4 + \dfrac{m_\Delta^2}{2} \text{Tr}\left[ \Delta^\dagger \Delta \right] + \dfrac{\lambda_{\Delta}}{4} \text{Tr}\left[\Delta^\dagger \Delta \Delta^\dagger \Delta \right]+ \dfrac{\lambda^{\prime}_{\Delta}}{4} \text{Tr}\left[\Delta^\dagger \Delta \right]^2 \nonumber \\ & + \dfrac{m_\Omega^2}{4}\text{Tr}\left[\Omega^\dagger \Omega\right] + \dfrac{\lambda_{\Omega}}{16} \text{Tr}\left[\Omega^\dagger \Omega \right]^2+  \dfrac{1}{8} \lambda_{\Delta \Omega} \text{Tr}\left[\Delta^\dagger \Delta\right]  \text{Tr}\left[\Omega^\dagger \Omega\right] \nonumber \\ & +\dfrac{1}{2} \lambda_{H\Delta} H^\dagger \Delta \Delta^\dagger H + \dfrac{1}{2} \lambda^{\prime}_{H\Delta} \text{Tr}\left[\Delta^\dagger \Delta\right] H^\dagger H +\dfrac{1}{2}\lambda_{H\Omega} H^\dagger \Omega \Omega^\dagger H \nonumber \\ & + \dfrac{1}{4} s_{\kappa} \kappa \left( H^T \widetilde{\Delta} \Omega H +\text{h.c.}\right) \, ,
\end{align}
where $s_\kappa=\pm1$ is the sign of the dimensionless coupling $\kappa$ chosen to be positive. All $\lambda^{(\prime)}_{i}$ and $m_i$  couplings with $i=\{h,\Delta,\Omega,\Delta \Omega, H \Delta, H\Omega \}$ are free parameters and have respectively dimension zero and dimension one.
We used the symbol $\widetilde{\Delta} \equiv \left(i\sigma_2 \right) \Delta^\dagger$ and in the following we parametrize the Higgs doublet as $H=(v_h+h)/\sqrt{2}$ in unitary gauge. $v_h$ is the vev related to the couplings of the scalar potential $\lambda_h = m_h^2 / (2 v_h^2)$, and $\mu_h^2 = m_h^2/2$. $h$ denotes the real Higgs scalar degree of freedom. Our model provides a scalar that fits the version in the SM Higgs boson with $m_h = 125.10 \pm 0.14$~GeV and $v_h = 246.22$~GeV~\cite{Zyla:2020zbs}.

\subsubsection{Masses and mixings}

The electroweak spontaneous symmetry breaking induces extra terms, proportional to the Higgs vev $v_h$ and induced by the $\kappa$-coupling, in the scalar mass-matrix which result in a mixing between scalars with identical quantum numbers, namely 2 CP even neutral scalars and 2 charged scalars with $Q=1$. In the following, we explicit the mass eigenstates and eigenvalues for these scalars as well as the masses of the neutral CP-odd and $Q=2$ scalars.\par \medskip

\paragraph{Neutral CP-even scalars:}
After electroweak symmetry breaking, the mass matrix for the CP-even neutral scalars in the basis $(\Omega^0_R,\Delta^0_R)$ reads
\begin{equation}
\label{eq:neutral_massmatrix}
\mathcal{M}^2_0=\left(
\begin{array}{cc}
 \dfrac{1}{2} v_h^2 \lambda_{H\Omega}+m_\Omega^2 & \dfrac{s_\kappa v_h^2 \kappa }{4} \\
 \dfrac{s_\kappa v_h^2 \kappa }{4} &\dfrac{1}{2} v_h^2 \left(\lambda_{H\Delta} +\lambda_{H\Delta}^\prime \right)+m_{\Delta}^2 \\
\end{array}
\right) \, ,
\end{equation}
that can be diagonalized by the rotation 
\begin{equation}
    \left(
\begin{array}{c}
 \Omega_R^0 \\
 \Delta^0_R \\
\end{array}
\right)= R_0 \left(
\begin{array}{c}
 S^0_1 \\
 S^0_2 \\
\end{array}
\right) \, \equiv \, \left(
\begin{array}{cc}
 \cos \theta_0 &  \sin \theta_0 \\
  -\sin \theta_0 &  \cos \theta_0  \\
\end{array}
\right) \left(
\begin{array}{c}
 S^0_1 \\
 S^0_2 \\
\end{array}
\right)  \, ,
\end{equation}
where $R_0$ is a rotation matrix, and $S_1^0$ and $S_2^0$ are the mass eigenstates with corresponding masses
\begin{eqnarray}
m_{S^0_{1,2}}^2& = &\dfrac{1}{4}  \Big(  v_h^2 \left( \lambda_{H\Delta}+\lambda_{H\Omega}+ \lambda^\prime_{H\Delta}\right)+2 \left(m_{\Delta }^2+m_\Omega^2\right)   \\ 
& &\pm \sqrt{\Big(v_h^2 \left( \lambda_{H\Delta}-\lambda_{H \Omega}+ \lambda^\prime_{H\Delta}\right)+2 \left(m_{\Delta }^2-m_{\Omega }^2\right)\Big)^2+ \kappa ^2 v_h^4}\, \Big)\,,
\label{eq:mass_splitting_neutral-scalars}
\end{eqnarray}
where the $-$ sign applies to $m_{S^0_{1}}$ and the $+$ sign to $m_{S^0_{2}}$. Further details on the mixing angle can be found in Appendix~\ref{sec:rotation}. \par \medskip

\paragraph{Neutral CP-odd scalar:}
Only one CP-odd scalar is present in the spectrum, which is denoted by $\tilde{S}^0$ in the following. This particle corresponds to the imaginary part of $\Delta^0$, $\tilde{S}^0 \equiv \Delta_I^0$. The mass of this scalar is given by
\begin{equation}
    m_{\tilde{S}^0}^2=\dfrac{1}{2}  v_h^2(  \lambda_{H\Delta}+\lambda^\prime_{H\Delta} )+m_{\Delta }^2\,,
\end{equation} \par \medskip 
\paragraph{Charged scalars with $Q=1$:}
The mass matrix for charged scalars in the basis $(\Omega^\pm,\Delta^\pm)$ reads
\begin{equation}
\label{eq:charged_massmatrix}
\mathcal{M}^2_\pm=\left(
\begin{array}{cc}
\dfrac{1}{2} v_h^2 \lambda_{H\Omega}+m_{\Omega}^2  &- \dfrac{ s_\kappa v_h^2 \kappa }{4 \sqrt{2}} \\
-  \dfrac{ s_\kappa v_h^2 \kappa }{4 \sqrt{2}} &\dfrac{1}{4} v_h^2 \left(\lambda_{H\Delta} +2\lambda_{H\Delta}^\prime \right)+m_{\Delta}^2 \\ 
\end{array}
\right)\, ,
\end{equation}
that can be diagonalized via a rotation matrix connecting the gauge and mass basis by the transformation: 
\begin{equation}
    \left(
\begin{array}{c}
 \Omega^\pm \\
 \Delta^\pm \\
\end{array}
\right)=  R_1 \left(
\begin{array}{c}
 S^\pm_1 \\
 S^\pm_2 \\
\end{array}
\right) \, ,
\end{equation}
where $S_1^\pm$ and $S_2^\pm$ are the mass eigenstates with masses
\begin{align}
    m_{S^\pm_{1,2}}^2=\dfrac{1}{8}  \Big( &  v_h^2 \left( \lambda_{H\Delta}+2\lambda_{H\Omega}+2 \lambda^\prime_{H\Delta}\right)+4 \left(m_{\Delta }^2+m_\Omega^2\right) \nonumber \\ & \left. \pm \sqrt{\Big(v_h^2 \left( \lambda_{H\Delta}-2\lambda_{H \Omega}+2 \lambda^\prime_{H\Delta}\right)+4 \left(m_{\Delta }^2-m_{\Omega }^2\right)\Big)^2+2 \kappa ^2 v_h^4}\right)\, ,
\end{align}
where the $-$ sign applies to $m_{S^\pm_{1}}$ and the $+$ sign to $m_{S^\pm_{2}}$. Further details on the mixing angle can be found in Appendix~\ref{sec:rotation}. \par \medskip

\paragraph{Charged scalar with $Q=2$}
One charged scalar with $Q=2$ is present in the spectrum. This particle corresponds to the state $\Delta^{++}$ and will be denoted in the following by $S^{\pm\pm}$ for notation consistency, with a corresponding mass given by
\begin{equation}
    m^2_{S^{\pm\pm}}=\dfrac{v_h^2 \lambda_{H\Delta}^\prime}{2}+m_\Delta^2\,,
\end{equation}

\subsubsection{Boundedness from below conditions for the scalar potential}
\label{sec:boundedfrombelow}

In order to ensure stability in the scalar sector, the scalar potential in Eq.~(\ref{eq:scalarpotential}) must satisfy the so-called ``boundedness from below" (BFB) conditions. Such conditions can be formalized as relations among the couplings in such a way that for a scalar potential with $N$ complex scalar fields $\phi_i$ with $i=1,..,N$: 
\begin{equation}
    \exists V_{\rm min} \in \mathbb{R}: \, V(\phi_1, \ldots, \phi_N) \ge V_{\rm min} \qquad \forall \, \phi_j \in \mathbb{C} \land \, 1\le j \le N \, ,
\end{equation}
where $V_{\rm min}$ is the potential minimum (see~\cite{Chakrabortty:2013mha, Kannike:2012pe, Ivanov:2018jmz} for further details).
For a renormalizable theory, the scalar potential is a polynomial function of the fields. Therefore, the condition is translated to 
\begin{equation}
    V(\phi_1, \ldots,\phi_N) \rightarrow \infty \quad {\rm for \, any }\, |\phi_j| \rightarrow \infty \, .
\end{equation}
The shape of the potential is controlled by the couplings, this sets relations between various couplings in order to have a correct vacuum and to avoid the transitions to "unbounded" vaccua. The scalar potential in Eq.~(\ref{eq:scalarpotential}) can be written in terms of the mass eigenstates, in such a way that it has the functional form:
\begin{equation}
    V = V_{\rm min} + M^2_{ij} \Phi_i \Phi_j + \omega_{ijk} \Phi_i \Phi_j \Phi_k + \lambda_{ijkl} \Phi_i \Phi_j \Phi_k \Phi_l \, ,
\end{equation}
where $\Phi_i$ is an array containing the scalar fields of the model both neutral and charged. We perform a partial analysis in the gauge basis of the fields.
Many of the terms within $M^2_{ij}$, $\omega_{ijk}$, and $\lambda_{ijkl}$ actually vanish due to gauge invariance and symmetry conditions. In our case, we focus on the quartic-terms as they govern the behaviour of the potential when $|\Phi_i| \rightarrow \infty$. To obtain the BFB conditions, we compute the fourth derivatives for every combination of the fields to obtain the couplings:
\begin{eqnarray}
    \dfrac{1}{\xi(i,j,k,l)} \frac{\partial^4 V}{{\partial \Phi_i}{\partial \Phi_j}{\partial \Phi_k}{\partial \Phi_l}} = \lambda_{ijkl} \, ,
\end{eqnarray}
where $\xi$ is the normalization factor:
\begin{align}
    \xi(i,j,l,k) = & 12 (\delta _{ij} \delta _{ik} \delta_{jl} \delta_{kl}) + 2 ( \delta_{ij} + \delta _{ik} + \delta_{il} + \delta_{jk} + \delta_{jl} + \delta_{kl}) \nonumber \\
    & + (1-\delta_{ij}) (1-\delta_{ik}) (1-\delta_{il}) (1-\delta_{jk}) (1-\delta_{jl}) (1-\delta _{kl}) \, ,
\end{align}
needed to take into account the numerical factors from the derivatives.

We classify the couplings depending of their parity under the sign change of any of the fields. Using this criterion, the couplings $\lambda_{iiii}$ and $\lambda_{iijj} (i\neq j)$  are even, and the rest of the couplings:  $\lambda_{iijk}$, $\lambda_{iiij}$, and $\lambda_{ijkl}$ are odd for $i\neq j \neq k \neq l$. In general terms, all couplings are constrained by perturbatibility: $0 \leq |\lambda_{ijkl}| <\sqrt{4\pi}$. However, the $\lambda_{iiii}$ must be always positive. The remaining couplings might be either positive or negative but restricted by the value of $\lambda_{iiii}$.

In our case, many of the conditions are either trivial or redundant with other relations. Nevertheless, the analysis of the fourth derivative presents a minimal approach to the BFB conditions. Due to the complexity of the potential, there might be non-trivial correlations between fields going to infinity that can lead to an unbounded vaccua. The relations coming from the $\lambda_{iiii}$ are the following:
\begin{align}
    \lambda_{\Delta} + \lambda_{\Delta}^\prime >0 \, , \qquad
    \dfrac{\lambda_{\Delta}}{2} + \lambda_{\Delta^\prime} >0 \, , \qquad
    \dfrac{\lambda_{\Omega}}{4} >0 \, , \qquad
    \lambda_h >0 \, ,
\end{align}
and the minimal copositivity criteria $\lambda_{iijj} + \sqrt{\lambda_{iiii}\lambda_{jjjj}} \geq 0$ leads to:
\begin{align}
    2(\lambda_{\Delta} + \lambda_{\Delta}^\prime) + \sqrt{\left(\dfrac{\lambda_{\Delta}}{2} + \lambda_{\Delta}^\prime \right)\left(\lambda_{\Delta} + \lambda_{\Delta}^\prime \right)} \geq 0 \, , \quad \quad
    \lambda_{\Delta\Omega} + \sqrt{\lambda_{\Omega} (\lambda_{\Delta} + \lambda_{\Delta}^\prime)} \geq 0 \, , \qquad \nonumber\\
    \lambda_{H\Delta} + \lambda_{H\Delta}^\prime + \sqrt{(\lambda_{1\Delta} + \lambda_{\Delta}^\prime)\lambda_H} \geq 0 \, , \quad
    \lambda_{H\Delta}^\prime + \sqrt{(\lambda_{\Delta} + \lambda_{\Delta}^\prime)\lambda_H} \geq 0\, , \quad \lambda_{H\Omega} + \sqrt{\lambda_{\Omega} \lambda_H} \geq 0 \, , \nonumber \\
    \lambda_{\Delta} + 3 \lambda_{\Delta}^\prime \geq 0 \, ,  \quad
    \lambda_{\Delta\Omega} + \sqrt{\lambda_{\Omega} \left(\dfrac{\lambda_{\Delta}}{2} + \lambda_{\Delta}^\prime \right)} \geq 0 \, , \quad
    \dfrac{\lambda_{H\Delta}}{2} + \lambda_{H\Delta}^\prime + \sqrt{\lambda_h \left(\dfrac{\lambda_{\Delta}^\prime}{2} + \lambda_{\Delta}^\prime\right)} \geq 0 \, .        
\end{align}

\subsection{Dark matter candidates}
As the $\mathbb{Z}_2$ discrete symmetry is unbroken in this model, any interaction process must feature an even number of $\mathbb{Z}_2$-odd particle. As a result, the lightest neutral $\mathbb{Z}_2$-odd particle of this model is stable, electrically neutral and features weak interactions, implying that it is a viable WIMP-like dark matter candidate. In our model, there are four neutral particles: 3 scalars $S_1^0,\, S_2^0,\, \tilde S^0$ and one fermionic state $f^0$ in the mass eigenstate basis. However, the mass of the state $S_2^0$ is larger than $S_1^0$ by definition of the mass eigenstates and eigenvalues in Eq.~(\ref{eq:mass_splitting_neutral-scalars}). In addition, one can show that the pseudoscalar $\tilde S^0$ mass is larger than the mass of $S_1^0$. This statement is not very obvious by comparing respective expressions for the masses of these states. However, using the alternative parametrization described in Appendix~\ref{sec:scan_strategy}, Eq.~(\ref{eq:mass_scalars_scan_parametrization}) makes this statement more explicit. As a result, we are left with two viable dark matter candidates, the $S_1^0$ scalar and the neutral $f^0$ fermion.

%\section{Neutrino masses and dark matter phenomenology}
\section{Phenomenological implications: neutrinos, dark matter, and lepton flavour violation}
\label{sec:neutrinodarkmatter}
In this section we provide details regarding the neutrino masses and dark matter phenomenology. We provide analytical expression of the neutrino masses in Sec.~\ref{sec:numasses}. In the following subsections, we discuss the dark matter relic abundance as well as dark matter direct and indirect detection. The complete numerical analysis of the parameter space is postponed to Sec.~\ref{sec:numerical_analysis}.

\subsection{Neutrino Masses}
\label{sec:numasses}
\begin{figure}[t!]
	\centering
	\includegraphics[width=0.5\columnwidth]{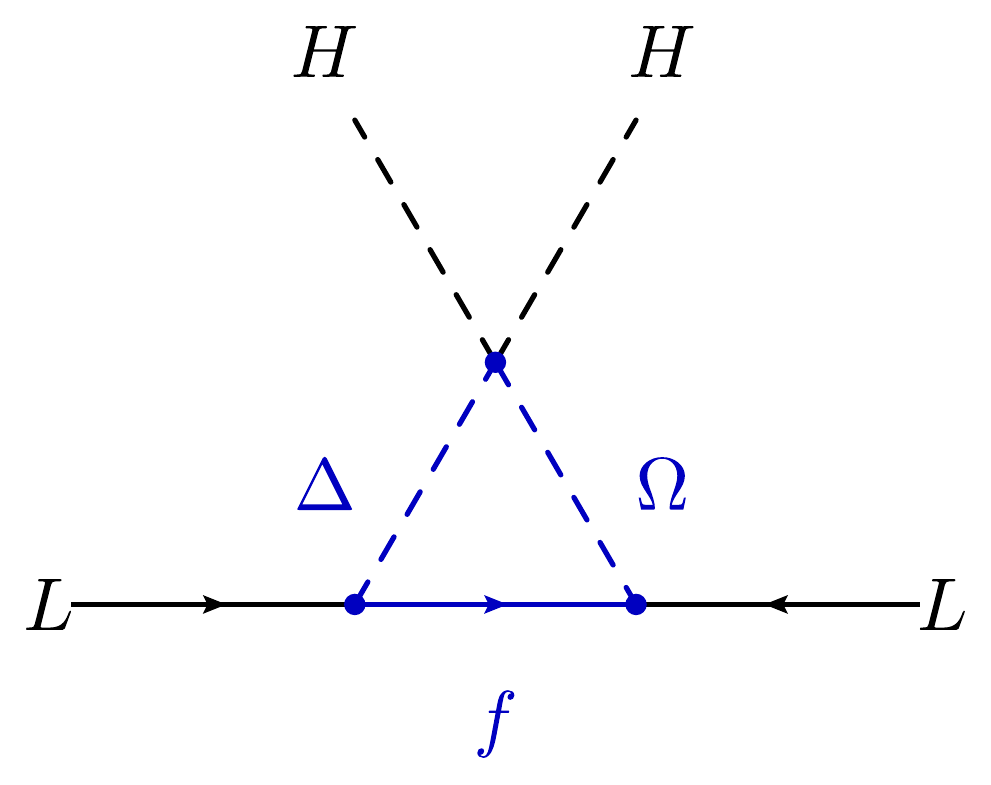}
	\caption{\label{fig:diagram} Diagram responsible for the one-loop generation of neutrino masses.}
\end{figure}

As advanced, neutrino masses are generated in the scotogenic model at the loop level by fields that are odd under the $\mathbb{Z}_2$ symmetry. Such one loop contributions are generated by self-energy diagrams involving loops of neutral fields $f^0, S^{0}_{1,2}$, as depicted in Fig.~\ref{fig:diagram} in the flavour-eigenstates basis. In our particular case, additional charged-scalar loops
exist~\cite{Hirsch:2000ef} generated by $f^\pm ,S^{\pm}_{1,2}$. Using the $\overline{\rm DR}$ scheme, the light neutrino masses are finite and can be expressed as
\begin{equation}
    m_{ij} = \frac{1}{16 \sqrt{2} \pi^2} \left( y_{\Delta}^i y_{\Omega}^j + y_{\Omega}^i y_{\Delta}^j\right) m_f \, F_{\rm loop} (m_{S^0_{1,2}},m_{S^\pm_{1,2}},m_f)\,,
\end{equation}
where the loop function is given by
\begin{align}
    F_{\rm loop}(m_{S^0_{1,2}},&m_{S^\pm_{1,2}},m_f)\,= \nonumber \\ \,&  \sin(2 \theta_{0}) \left(\frac{m^2_{S^0_1}}{m_f^2 - m^2_{S^0_1}} \log{\left(\frac{m^2_{S_1^0}}{m_f^2} \right)} - \frac{m^2_{S^0_2}}{m_f^2 - m^2_{S^0_2}} \log{\left(\frac{m^2_{S_2^0}}{m_f^2} \right)}\right) \nonumber \\
    & - \sin(2 \theta_1) \left(\frac{m^2_{S^\pm_1}}{m_f^2 - m^2_{S^\pm_1}} \log{\left(\frac{m^2_{S_1^\pm}}{m_f^2} \right)} - \frac{m^2_{S^\pm_2}}{m_f^2 - m^2_{S^\pm_2}} \log{\left(\frac{m^2_{S_2^\pm}}{m_f^2} \right)}\right)\,,
\end{align}
Therefore, neutrino masses are essentially controlled by the Yukawa couplings $y_\Omega^i$, $y_\Delta^i$, the coupling $\kappa$, via the mixing angles, which plays the role of a lepton number violating parameter, in addition to masses of the scalar states and new fermions. In the following, for simplicity, we assume a normal ordering (NO) for the neutrino masses. In this case the latest determination of the neutrino-sector parameters are given by Ref.~\cite{deSalas:2020pgw}, whose best-fit values and $\pm1\sigma$ range are 
\begin{align}
    \Delta m_{32}^2\,=\,  2.56^{+0.03}_{-0.04}\times 10^{-3}~\text{eV}^2 \, ,  \qquad \qquad \Delta m_{21}^2\,=\, 7.50^{+0.22}_{-0.20}\times 10^{-5}~\text{eV}^2 \, , \nonumber \\ 
    \theta_{12} \,=\,34.3^{+1.00}_{-1.00}\,^{\circ}  \, , \qquad  \theta_{23} \,=\,48.79^{+0.93}_{-1.25} \,^{\circ} \, , \qquad  \theta_{13} \,=\,8.58^{+0.11}_{-0.15}\,^{\circ} \, , \qquad \delta = 216^{+41}_{-25} \,{}^{\circ}.
    \label{eq:neutrino_oscillation_parameters}
\end{align}
In general terms, the Yuwaka sector ($y_\Omega^i$, $y_\Delta^j$) can contain 6 complex phases. Among them, 5 phases can be absorbed by field redefinitions and the remaining phase can be related to $\delta$. The impact of the CP-phase in our context is reduced due to the insensitivity of DM observables to neutrino flavours. For simplicity, as mentioned previously, we assume CP conservation therefore we set the usual Dirac phase $\delta$ to $0$. CP-violation could give rise to a contribution to the electron dipole moment at the two-loop level which is constrained at the level of $|d_e|/e<1.1\times10^{-29}\;\text{cm}$ \cite{Andreev:2018ayy}. Such effects have been studied in a similar context \cite{Abada:2018zra,Fujiwara:2020unw} and have been shown not to reduce significantly the available parameter space compatible with constraints from lepton-flavour violating observables. These effects are not expected to differ significantly in our setup.

In the following, we impose the neutrino-sector parameters of our model to reproduce the best fit values of Eq.~(\ref{eq:neutrino_oscillation_parameters}) from Ref.~\cite{deSalas:2020pgw}. In addition, we consider the lightest neutrino mass-eigenstate $\nu_1$ as massless. Assuming this hierarchy reduces significantly the number of the free Yukawa couplings of the model to just one. Details regarding the parametrization used to reproduced parameters from  Eq.~(\ref{eq:neutrino_oscillation_parameters}) can be found in Appendix~\ref{sec:appendix_neutrino_masses}.

\subsection{Dark matter direct detection}
\label{sec:DD}
Both fermionic and scalar dark matter candidates possess electroweak charges and significant interaction with light quarks therefore could potentially be observed by scattering on some detector material nuclei. In the following we evaluate the DM-nuclei scattering cross section for our potential fermionic and scalar dark matter candidates and discuss the expected values in light of current bounds from Direct Detection (DD) experiments.

\subsubsection{Fermionic dark matter}
In the case where the lightest neutral stable particle is the fermionic state $f^0$, a vector coupling exists between this state and the $Z-$boson of the SM allowing for a Spin-Independent (SI) cross section at tree-level. The amplitude for the scattering of our DM state and first-generation quark is determined by the charge assignment of these particles with respect to the electroweak symmetry group, the only free parameter being the DM mass. The corresponding expression for the DD cross section with a proton can be straightforwardly expressed as~\cite{Cirelli:2005uq}
\begin{equation}
    \sigma^p \,=\,  \frac{ G_F^2 m_p^2}{8 \pi} \Big(1-4s_W^2
    \Big)^2  \, \simeq 4.2 \times 10^{-41}\,\text{cm}^2 ,
\end{equation}
where $G_F$ is the Fermi constant, $s_W$ is the sine of the Weinberg angle, in the limit where the DM mass is much larger than the proton mass $m_f\gg m_p$. As long as this hierarchy is satisfied, this expression is independent of any parameter of the model and fixed. The most stringent constraints on the SI cross section are currently achieved by the Xenon1T experiment~\cite{Aprile:2018dbl} which excludes $\sigma^p\gtrsim5 \times 10^{-46}\,\text{cm}^2$ for a $50$ GeV DM mass and up to  $\sigma^p\gtrsim 10^{-44}\,\text{cm}^2$ for $10$ TeV DM masses. As a result, by requiring the DM mass to remain below the perturbative unitarity limit $\mathcal{O}(10-100)$ TeV, the fermionic DM candidate is entirely ruled out by the current constraints on SI cross section from direct detection.

\subsubsection{Scalar dark matter}
In the case where our DM candidate is the lightest neutral scalar $S^0_1$, a SI contribution to the direct detection cross section is generated at tree-level by the coupling of $S^0_1$ to the Higgs boson
\begin{equation}
    \mathcal{L}\, \supset \, - \dfrac{v_h}{2} h (S_1^0)^2 \Big(
  s_{\theta_0}^2\big( \lambda_{H \Delta}+ \lambda_{H \Delta}^\prime \big) +   c_{\theta_0}^2 \lambda_{H \Omega} + s_\kappa \kappa c_{\theta_0} s_{\theta_0} 
    \Big)  \, ,
\end{equation}
with $s_{\theta_{0}} \equiv \sin \theta_{0}$ and $c_{\theta_{0}} \equiv \cos \theta_{0}$. This induces an effective coupling to nucleons ($N$) of the form
\begin{equation}
  \mathcal{L}_\text{eff}^N \, = \, \lambda^\text{eff}_N  \bar N N (S_0^1)^2    \, ,
\end{equation}
where the effective coupling $ \lambda^\text{eff}_N  $ can be expressed as
\begin{equation}
\lambda^\text{eff}_N \, = \, f_N \left( \frac{m_N}{m_h^2} \right) \Big(
  s_{\theta_0}^2\big( \lambda_{H \Delta}+ \lambda_{H \Delta}^\prime \big) +   c_{\theta_0}^2 \lambda_{H \Omega} + s_\kappa \kappa c_{\theta_0} s_{\theta_0}
    \Big)  \, ,
\end{equation}
with $  f_N \, = \frac{2}{9} + \frac{7}{9} \sum_{q=u,d,s}f^{(N)}_{Tq} \, \approx 0.3$ being the scalar form factor. The SI DM-proton cross section can be straightforwardly derived: 
\begin{align}
\sigma^p_\text{tree} \,=\,  \frac{ \mu^2 m_p^2 f_N^2}{4 \pi m_{S_1^0}^2 m_h^4} \Big(
  s_{\theta_0}^2\big( \lambda_{H \Delta}+ \lambda_{H \Delta}^\prime \big) +   c_{\theta_0}^2 \lambda_{H \Omega} + s_\kappa \kappa c_{\theta_0} s_{\theta_0} 
    \Big)^2  \,  .
    \label{eq:sigmaDD_tree}
\end{align}
We checked that this expression is in agreement with Ref.~\cite{Casas:2017jjg} and Ref.~\cite{Chao:2018xwz} in the limit of vanishing mixing angles and agrees numerically with results from the \texttt{micrOMEGAS} code~\cite{Belanger:2001fz,Belanger:2018ccd}.\par\medskip

However, electroweakly charged DM candidates can have sizable loop-contributions to direct detection cross section induced by box diagrams involving electroweak gauge fields~\cite{Cirelli:2005uq}. At the loop-level, the neutral component of a hypercharge $Y$ scalar triplet interacts with quarks via a twist-2 operator in the large-mass expansion as~\cite{Drees:1993bu,Hisano:2011cs,Chao:2018xwz} 
\begin{equation}
    \mathcal{L}_\text{eff}^q \, \supset \, \dfrac{f_T^Y}{m_{S_1^0}^2} S_1^0 (i\partial^\mu) (i\partial^\nu)  S_1^0 \mathcal{O}_{\mu \nu }^q \,,
\end{equation}
where
\begin{equation}
    \mathcal{O}_{\mu \nu }^q \, \equiv \, \dfrac{1}{2} \bar q i \Big( D_\mu \gamma_\nu+D_\nu \gamma_\mu-\dfrac{1}{2} g_{\mu \nu} \slashed{D} \Big)\,.
\end{equation}
We use the parametrization from~\cite{Chao:2018xwz}
\begin{equation}
    f_T^Y\, = \, \dfrac{\alpha_2^2}{8 m_W^2} \big(2-Y^2 \big) \, F \left(\dfrac{m_W^2}{m_{S_1^0}^2}\right) \,,
\end{equation}
with
\begin{equation}
    F(r)\,\equiv \, \left[ r \log r + 4 + \dfrac{(4-r)(2+r)}{\sqrt{1-r/4}\sqrt{r}} \arctan \left( \dfrac{2 \sqrt{1-r/4}}{\sqrt{r}}\right) \right] \,,
\end{equation}
which reduces to $ F(r) \simeq 4 \pi /\sqrt{r}$ in the limit $r \ll 1$. In our case the DM candidate is a mix of neutral scalar components of $Y=0$ and $Y=1$ triplets, and several quasi-degenerate charged and neutral scalar states are present in the spectrum. The precise estimate of the one-loop contribution to the direct detection cross section would require a rather cumbersome and dedicated analysis which goes beyond the scope of this paper. Therefore, in the limit where the mass splitting between the various scalar components is small -- assumption justified as illustrated in Appendix~\ref{sec:additional_plots} -- we can estimate the one-loop amplitude as the sum of the contributions from $Y=0$ and $Y=1$ triplets weighted respectively by $\cos^2(\theta_0)$ and $\sin^2(\theta_0)$, which should reproduce the expression from Ref.~\cite{Chao:2018xwz} in the limit of vanishing mixing angles. This gives the following estimate of the cross section by including the electroweak one-loop corrections:
\begin{align}
    \sigma^p_{\rm tree+loop} \,=\,&  \frac{ \mu^2 m_p^2}{ 4\pi m_{S_1^0}^2} \left[ \dfrac{3}{4} c_{\theta_0}^2f_T^{0} f_p^{\rm PDF} + \dfrac{3}{4} s_{\theta_0}^2 f_T^{1} f_p^{\rm PDF} \right. \nonumber \\ & + \left. \dfrac{f_N^2}{m_h^4}\Big(
  s_{\theta_0}^2\big( \lambda_{H \Delta}+ \lambda_{H \Delta}^\prime \big) +   c_{\theta_0}^2 \lambda_{H \Omega} + s_\kappa \kappa c_{\theta_0} s_{\theta_0} 
    \Big) \right]^2\,,
    \label{eq:sigmaDD_tot}
\end{align}
where $f_p^{\rm PDF}=0.526$~\cite{Hisano:2015rsa} is the second moment of proton parton distribution function (PDF) evaluated at the scale $\mu = m_Z$. In this case, the SI cross section depends on numerous parameters of the model and the numerical evaluation of this quantity is investigated in details further on in Sec.~\ref{sec:directdetectionnum}.

\subsection{Dark matter relic abundance}
In scotogenic models, the dark matter relic abundance can be generated by the usual freeze-out mechanism. Indeed, the electroweak charge of our DM candidate ensures thermalization with the SM bath until the DM particles become non-relativistic and annihilate. In the limit where our DM candidate is $\Omega_R^0$-like (the neutral component of the $Y=0$ triplet), a large contribution to the DM annihilation cross section, induced by gauge interactions, features weak bosons in the final state such as $W^+ W^-$. Such contribution can be expressed as
\begin{equation}
    \langle \sigma v \rangle_{W^+ W^-} \, \simeq\, 4 \times 10^{-26}\, \text{cm}^3 \, \text{s}^{-1} \left( \dfrac{1.2 ~\text{TeV}}{m_{S^0_1}}\right)^2\, ,
\end{equation}
which essentially depends only on the DM mass. This sets the typical mass scale for our DM candidate. For DM particles with masses lighter than $\sim 1.2$ TeV, processes of the kind $\text{DM}+\text{DM} \rightarrow \text{SM}+\text{SM}$ are too efficient resulting in a density to small to account for all the observed DM relic abundance and must rely on additional processes. For masses above $\sim 1.2$ TeV, the observed DM relic abundance can be achieved by DM annihilations to electroweak SM bosons via gauge interactions, i.e. $W^+W^-$, $Z Z$. In addition, annihilations to Higgs bosons pair $hh$ via a combination of couplings $\kappa$, $\lambda_{H \Omega}$, $\lambda_{H \Delta}$ and $\lambda_{H \Delta}^\prime$ depend on the specific mixing between the neutral components of the $Y=0$ and $Y=1$ new scalar multiplets. Annihilation cross sections to SM fermions, mediated by $s-$channel Higgs-boson, are typically helicity suppressed and do not contribute significantly, except for the top-quark case. However, annihilations into SM leptons via the Yukawa couplings $y_\Omega^i$ and $y_\Delta^i$, mediated by $t-$channel exchange of heavy fermionic states $f$, can significantly contribute. As mentioned in the previous section, only one of these Yukawa coupling is a free parameter as we imposed to reproduce observed parameter of the neutrino sector.\par\medskip
However, as the DM candidate is the lightest and only stable $\mathbb{Z}_2$-odd particle, any heavier $\mathbb{Z}_2$-odd state is unstable and will eventually decay to a DM particle and additional states. By conservation of number density in the  $\mathbb{Z}_2$-odd sector, the various new scalars (charged and neutrals) and new fermionic states will contribute to the co-annihilation cross section via $2 \rightarrow 2$ processes where the initial states are a pair of $\mathbb{Z}_2$-odd particles and the final states DM particles. In addition, as shown further on, the mass splitting between these states and the DM candidate is typically small implying that co-annihilations are sizable and have to be taken into account to estimate the DM relic density. The numerous diagrams involved in  (co-)annihilation processes are depicted in Fig.~\ref{fig:diagrams1} and Fig.~\ref{fig:diagrams2} for illustration. 
As the number of involved states and diagrams is rather large, the relic abundance has to be computed numerically. For this purpose, we rely on the code \texttt{micrOMEGAS}~\cite{Belanger:2001fz,Belanger:2018ccd} after implementing the model both in \texttt{LanHEP}~\cite{Semenov:2014rea} and \texttt{Feynrules}~\cite{Alloul:2013bka} as cross check. \par\medskip

\subsection{Indirect searches}
\label{sec:indirectsearches}

\begin{figure}[!t]
\centering
    \includegraphics[width=0.8\textwidth]{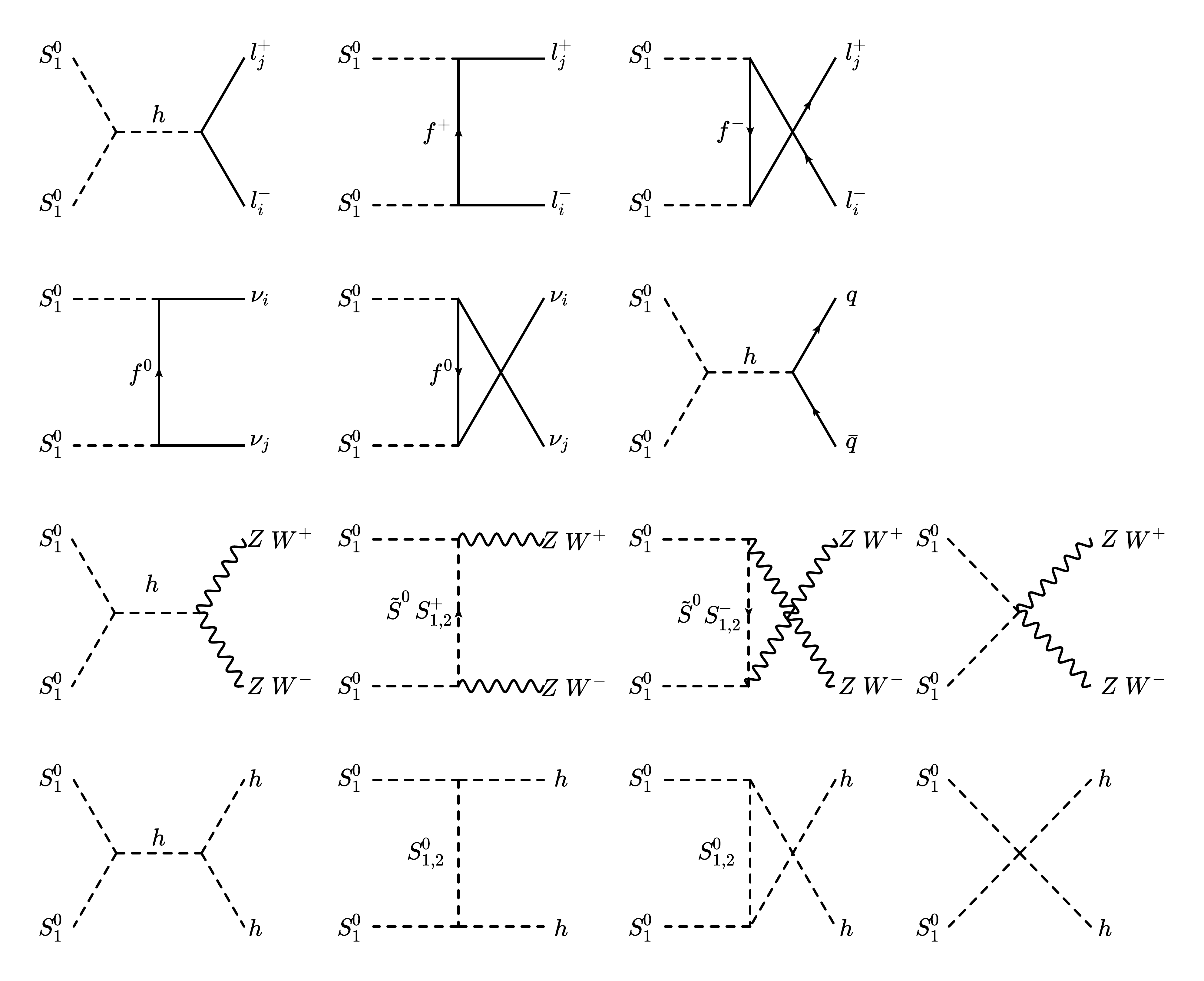}
    \caption{Dark Matter annihilations into SM particles.}
    \label{fig:indirect}
\end{figure}

Dark matter annihilations within the halo of the galactic center or in galactic subhalos might produce $\gamma$-ray signals potentially observable by the current and future generation of telescopes, depending on the annihilation channels. In order to illustrate the relevance of various annihilation channels, represented in Fig.~\ref{fig:indirect}, we provide analytical expressions for the most relevant velocity-averaged annihilation cross sections assuming our DM candidate to be $S_1^0 \simeq \Omega_R^0$, in the limit of vanishing mixing angles and assuming $m_{S^0_1} \simeq m_{S^1_1} \gg m_h, m_W, m_Z$. We performed a velocity expansion at leading order in the mean DM velocity $\langle v_{\rm DM} \rangle \ll 1$. In this limit, the DM annihilation cross section into a pair of gauge bosons is given by the $s$-wave terms
\begin{equation}
    \langle \sigma v \rangle_{W^+ W^-} \, \simeq\, \frac{e^4}{8 \pi m_{S^0_1}^2 s_W^4}+\frac{\lambda_{H \Omega}^2}{64 \pi  m_{S^0_1}^2} \,,
    \label{eq:sigmavWW}
\end{equation}
and
\begin{equation}
    \langle \sigma v \rangle_{ZZ} \, \simeq\, \dfrac{\lambda_{H \Omega}^2}{128 \pi  m_{S^0_1}^2} \,,
\end{equation}
with $e \equiv \sqrt{4 \pi \alpha_{\rm EM}}$. Additional $s-$wave terms induced by the couplings $\lambda_{H \Delta}, \lambda_{H \Delta}^\prime$ and $\kappa$ are also present but suppressed by mixing angles. Annihilation cross section into a pair of Higgs bosons is given by
\begin{align}
    \langle \sigma v \rangle_{hh}\, \simeq & \,\dfrac{1}{128 \pi  e^4 m_{S_1^0}^6 \big(m_{S_1^0}^2+m_{S_2^0}^2\big)^2} \left(e^2 \lambda_{H\Omega } m_{S_1^0}^2 \left(m_{S_1^0}^2+m_{S_2^0}^2\right) \right. \nonumber \\ & \left.- 2 m_W^2 s_W^2 \left(2 \lambda_{H\Omega}^2 \big(m_{S_1^0}^2+m_{S_2^0}^2\big)+\kappa^2 m_{S_1^0}^2 \right)\right)^2 \, .
\end{align}
Annihilations to a pair of charged leptons $\ell^+_i \ell^-_i$ correspond to the following cross section
\begin{equation}
  \langle \sigma v \rangle_{\ell^+_i \ell^-_i}\, \simeq \, \dfrac{(y_{\Omega }^i)^4 m_f^2 }{2 \pi  \big(m_f^2+m_{S_1^0}^2\big)^2}+\frac{\lambda_{H\Omega}^2 m_\ell^2}{128 \pi  m_{S_1^0}^4} \, ,
  \label{eq:sigmavLL}
\end{equation}
where $i=1,2,3$ is a flavour index. The second term being helicity suppressed, annihilations are mostly efficient for heavy leptons, i.e. $\tau^+ \tau^-$, while annihilations into a pair of neutrinos $ \nu_i \nu_i$ are only triggered by the Yukawa coupling $y_{\Omega }^i$ as
\begin{equation}
  \langle \sigma v \rangle_{ \nu_i \nu_i}\, \simeq \, \dfrac{(y_{\Omega }^i)^4 m_f^2 }{\pi  \big(m_f^2+m_{S_1^0}^2\big)^2}\,.
   \label{eq:sigmavnunu}
\end{equation}
For low masses $m_{S_1^0}\lesssim 3-4$ TeV, $W^+ W^-$ is the most promising channel as annihilations to this final state are not suppressed by any mixing angle for $S_1^0 \simeq \Omega_R^0$ and depends only on the DM mass. In addition, for large values of the couplings of the scalar potential, the $ZZ$ and $hh$ final state would become equally important. The HESS experiment is currently the most sensitive to annihilations for masses $m_{\rm DM}\gtrsim 100~\text{GeV}$. As no signal from DM annihilations has been detected so far, a constraint of the order of $\langle \sigma v \rangle \gtrsim 10^{-26}-10^{-25}~\text{cm}^3\,\text{s}^{-1}$ has been set by the HESS collaboration~\cite{Abdallah:2016ygi} for $\tau^+ \tau^-$ and  $W^+W^-$ final states. Moreover, CTA should be able to probe the vanilla value of the velocity averaged annihilation cross section $\langle \sigma v \rangle = 3 \times 10^{-26}$~\cite{CTAConsortium:2018tzg,Pierre:2014tra,Silverwood:2014yza,Lefranc:2015pza} for DM annihilations into specific channels such as $\bar b b$, $\tau^+ \tau^-$ and  $W^+W^-$. For larger masses  $m_{S_1^0}\gtrsim 3-4$ TeV, leptonic final state could dominate for sizable values of the Yukawa couplings. CTA sensitivity for DM annihilations to neutrinos should reach $\langle \sigma v \rangle_{\nu \nu} \sim 10^{-24}\,\text{cm}^3\,\text{s}^{-1}$ for $m_\text{DM}\simeq 1~\text{TeV}$~\cite{Queiroz:2016zwd}. However, the KM3Net experiment should reach the best sensitivity for such DM masses, by probing values as low as $\langle \sigma v \rangle_{\nu \nu} \gtrsim 10^{-25}\,\text{cm}^3\,\text{s}^{-1}$~\cite{2019ICRC...36..552G,Arguelles:2019ouk}. More details regarding our numerical results and the effect of such bounds are discussed in the following section.

%%%%%%%%%%%%%%%%%%%%%%%%%%%%%%%%%%%%%%%%%%%%%%%%%%%%%%%%%%%%%%%
\subsection{Lepton flavour violation}
\label{sec:LFV}

\begin{figure}[tb]
    \centering
    \includegraphics[width=1.0\columnwidth]{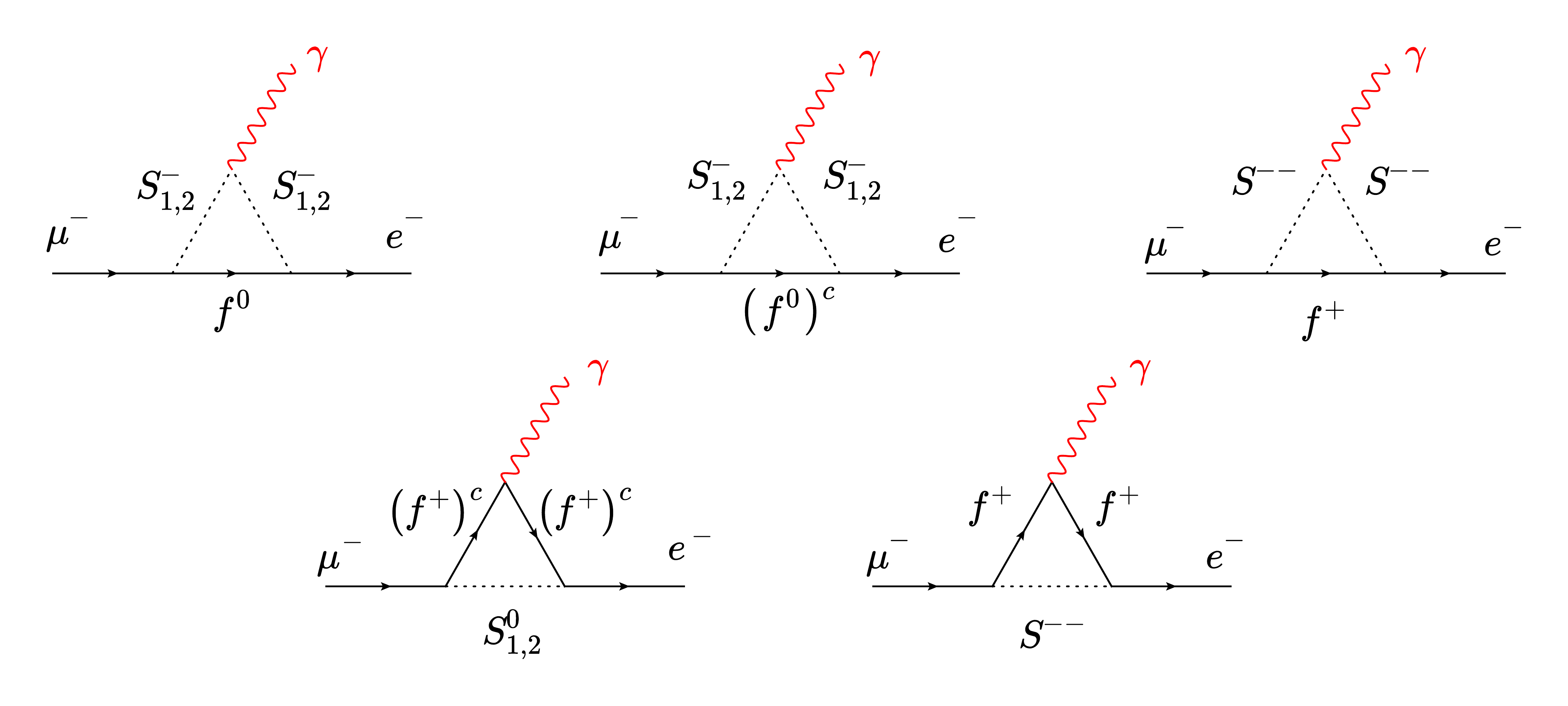}
    \caption{Diagrams contributing to $\mu \rightarrow e \gamma$. In total, there are 8 diagram with non-zero contribution to the process.}
    \label{fig:muegamma}
\end{figure}
One of the typical signatures of scotogenic-like models is rare lepton-flavour-violating leptonic decays~\cite{Toma:2013zsa,Akeroyd:2009nu} such as $\ell_\alpha \rightarrow \ell_\beta  \gamma$ or $\ell_\alpha \rightarrow \ell_\alpha \ell_\beta \ell_\beta$. One of the strongest bounds on such processes impose $\text{BR}(\mu^+ \rightarrow e^+ \gamma) < 4.2 \times 10^{-13}$~\cite{TheMEG:2016wtm} and $\text{BR}(\mu^+ \rightarrow e^+ e^+ e^-) < 1  \times 10^{-12}$~\cite{Bertl:1985mw}. In our setup,
$\ell_\alpha \rightarrow \ell_\beta \gamma$ is generated at the loop level by the following effective operator
\begin{equation}
    \mathcal{L}\,\supset \,\left(\dfrac{\mu_{\alpha \beta}}{2}\right)\bar \ell_\beta \sigma^{\mu \nu} \ell_\alpha F_{\mu \nu}\,.
\end{equation}
The branching fraction for $\mu \rightarrow e \gamma$ can be expressed as \cite{Toma:2013zsa}
\begin{equation}
    \text{BR}\big(\mu \rightarrow e \gamma\big)\,=\,\dfrac{3(4\pi)^2}{ G_F^2 m_\mu^2} \, \mu_{\mu e}^2 \,\text{BR}\big(\mu \rightarrow e \bar \nu_e  \nu_\mu \big) \,.
    \label{eq:expressionBRmutoegamma}
\end{equation}
The diagrams involved in the calculation of $\mu_{\mu e}$ are depicted in figure~\ref{fig:muegamma}. The complete expression for the coefficient $\mu_{\mu e}$, computed for our model, can be found in Eq. (\ref{eq:mutoegamma_mufactor}) of the appendices and depends on many various parameters of the model. In order to provide a simple numerical estimate, we consider the limit where the mass splittings between the DM candidate and the various scalars and vector-like fermion is small. In addition, by assuming universal Yukawa couplings $\bar y=y_{\Delta}^i=y_{\Omega}^j$ and vanishing mixing angles, the branching fraction is approximately given by
\begin{equation}
    \text{BR}\big(\mu \rightarrow e \gamma\big)\,\simeq\,\dfrac{10 e^2 \bar y^4 }{ 768 \pi^2 G_F^2 m_{S_1^0}^4 } \, \simeq 5.5 \times 10^{-11} \left( \dfrac{\bar y}{10^{-1}} \right)^4 \left( \dfrac{1\;\text{TeV}}{m_{S_1^0}} \right)^4\,,
\end{equation}
with $e=\sqrt{4\pi \alpha_{\rm em}}$. The constraint $ \text{BR}\big(\mu \rightarrow e \gamma\big)<4.2 \times 10^{-13}$ translates into a bound  
\begin{equation}
    \bar y \, < \, 3\times 10^{-2}\, \left( \dfrac{m_{S_1^0}}{1\;\text{TeV}} \right)\,.
    \label{eq:constraint_bary}
\end{equation}
One-loop contributions to the process $\mu \rightarrow 3e$ arise from box diagrams as well as $\gamma$, $Z$ and Higgs-penguin diagrams. The Higgs diagrams, involving suppressed yukawa couplings, are negligible for the first and second generation of leptons. The $Z-$penguins have been shown \cite{Toma:2013zsa} to be suppressed by the charged lepton masses and therefore are subdominant. In case where the dipole-like contribution to the photon penguin diagram is dominant, the decay rate for $\mu \rightarrow 3e$ becomes proportional to the $\mu \rightarrow e \gamma$ rate and is suppressed by an additional fine structure constant and phase space volume. \cite{Toma:2013zsa} have shown that non-dipole photonic diagrams never exceed the dipole contribution. As a result, only box diagrams can lead to a decay rate $\mu \rightarrow 3e$ larger than $\mu \rightarrow e \gamma$. The ratio of these two contributions would be suppressed by factor $\sim \bar y^4/(48 \pi^2 e^2)$ times a ratio of the corresponding loop functions. Since from Eq. (\ref{eq:constraint_bary}), the typical Yukawa coupling $\bar y$ is already constrained by $\mu \rightarrow e \gamma$ to be small, we expect the rate $\mu \rightarrow 3e$ to be subdominant therefore less constraining than  $\mu \rightarrow e \gamma$. This ratio has been computed explicitly in \cite{Toma:2013zsa}, for the normal hierarchy case, and is typically of order $\sim10^{-2}$ for $m_{\nu_1}=0$, assumption that we made throughout this work.

\section{Numerical analysis of the parameter space}
\label{sec:numerical_analysis}
\subsection{Scan of the parameter space}
We performed a scan in the parameter space and selected the points satisfying the relic density condition $\Omega_{\rm DM} h^2 \in [0.11933 \pm 3\times 0.00091]$ as a $3\sigma$ interval around the Planck best fit value $\Omega_{\rm DM} h^2 =0.11933$ for TT, TE, EE+lowE+lensing+BAO~\cite{Ade:2015xua}, computed numerically using the \texttt{micrOMEGAS}~\cite{Belanger:2001fz,Belanger:2018ccd} code. In addition, we imposed conditions allowing to reproduce the "normal-ordering" neutrino mass hierarchy as described in Sec.~\ref{sec:numasses} and lepton flavour violation constraints from $\mu \rightarrow e \gamma$ as described in Sec.~\ref{sec:LFV}. In order to perform the scan more efficiently, we defined convenient variables whose definitions can be found in Appendix~\ref{sec:scan_strategy}. In order to perform the scan, we generated random numbers in the following range of the various couplings
\begin{align}
    \sin (\theta_0)\, , \sin (\theta_1)\, \in \, & [-1,1] \, , \nonumber \\
    \lambda\, , \kappa\, , \hat{y}_\Delta \, \in \, & [10^{-10},\sqrt{4 \pi}]\, , \\
    m_{S_1^0} \, \in \, &[700,3\times 10^4 ]~\text{GeV}\, , \nonumber\\
    m_{f} \, \in \, &[700,3\times 10^5 ]~\text{GeV} \, .\nonumber
\end{align}
where $\hat{y}_\Delta$ is an effective Yukawa coupling relevant for the neutrino sector, as detailed in Appendix~\ref{sec:scan_strategy}. 
$\lambda$ denotes all the couplings labeled as $\lambda_i^{(\prime)}$ with $i=\{\Delta,\Omega,\Delta \Omega, H \Delta, H\Omega \}$. We take $\sqrt{4 \pi}$ as perturbative limit (upper bound) for the dimensionless couplings. In addition, we imposed a relative mass splitting  $(m-S^0_1)/S^0_1 >10^{-5}$ where $m$ denotes the mass of any $\mathbb{Z}_2$-odd state, to ensure efficient numerical convergence. We imposed the bounded-from-below conditions for the scalar potential, described in Sec.~\ref{sec:boundedfrombelow}. We considered the loop-induced mass splitting between the charged $f^+$ and neutral $f^0$ fermions as described in Sec.~\ref{sec:fermionsector}.
The scan is performed in log-space except for the couplings $\sin \theta_{0,1}$ where the scan is done in linear space on the variables $\psi_{0,1}$ defined in Appendix~\ref{sec:scan_strategy}. We split our scan in the parameter space in 2 regions, corresponding to $m_{S^0_1}>1.2$ TeV and $m_{S^0_1}<1.2$ TeV, where we run a longer scan in the former case as the relic density is much more sensitive to the various couplings of the models as in the later case.

\subsection{Neutrino sector}
\begin{figure}[!t]
\centering
    \includegraphics[width=0.6\textwidth]{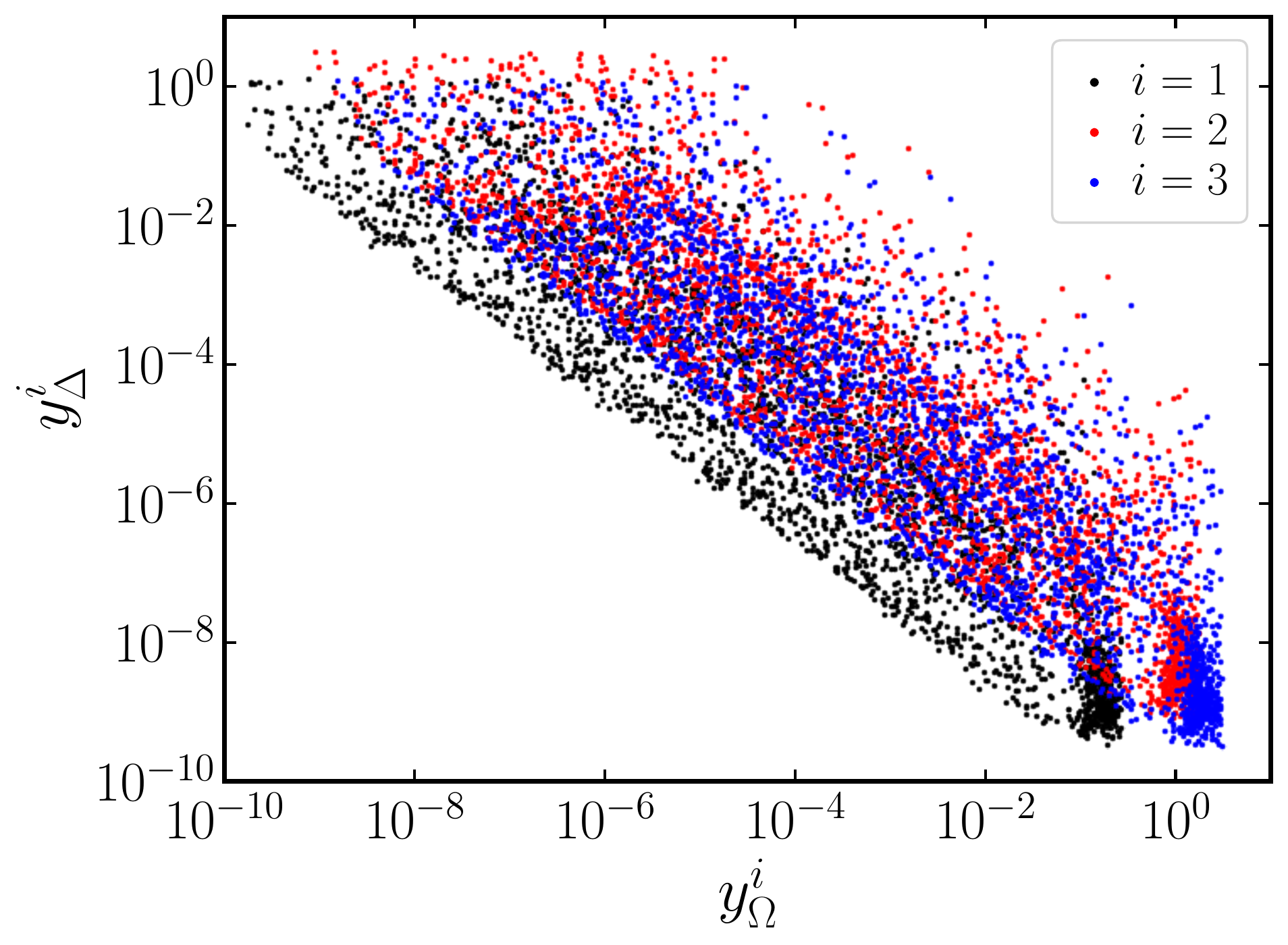}
    \caption{Scan in the parameter space allowing to reproduce the correct dark matter relic density as well as neutrino masses and mixings as described in Sec.~\ref{sec:numasses}. The results from the scan is represented in the plane $(y_\Omega^i,y_\Delta^i)$ where the 3 flavours indices $i=1,2,3$ are represented respectively in black, red and blue.}
    \label{fig:yukawa_Omega_Delta}
\end{figure}
The numerical results of our scan are represented in the plane $(y_\Omega^i,y_\Delta^i)$ in Fig.~\ref{fig:yukawa_Omega_Delta}.
As discussed in Sec.~\ref{sec:numasses}, the neutrino masses are proportional to the product of couplings $\propto y_\Omega \, y_\Delta$ therefore the parameter space shown in Fig.~\ref{fig:yukawa_Omega_Delta} corresponds to a "broad line" in log-log space roughly defined by $y_\Omega \, y_\Delta \sim 10^{-7}-10^{-10}$ depending on the flavour index. The broadness of the "line" -- for a fixed value of $y_\Delta^i$ -- corresponds to a variation of these Yukawa couplings compensated by other relevant couplings, controlling neutrino masses such as mixing angles and masses. The Yukawa couplings are strongly constrained by the neutrino masses but could be as low as $\sim 10^{-10}$. However they cannot be much lower as this would imply having Yukawa couplings that reach the perturbative limit that we imposed on our scan in the parameter space. The small values $y_\Delta, y_\Omega \sim 10^{-10}$ are theoretically not the most appealing part of our parameter space as it is not much more "natural" than adding right-handed neutrino singlets coupled to the Higgs and left-handed neutrinos. However a large $y_\Delta$ implied by a feeble value of $y_\Omega$ leads to interesting detection prospects as discussed further on. In addition, such large hierarchy $y_\Delta \sim 10^{\pm 10} y_\Omega$ is present typically for large DM masses $m_{S_1^0} \gtrsim 3-4 $ TeV, as illustrated in Fig.~\ref{fig:yukawacouplings}. Indeed in this part of the parameter space, the gauge contribution to the annihilation cross section is no longer efficient enough and in order to achieve the correct DM relic density, annihilations have to occur via the Yukawa couplings in addition to the various couplings of the scalar potential such as $\kappa$ for instance which is also constrained by the neutrino masses. Combining this effect with the perturbative limit on our parameters sets an upper bound on the DM mass of the order of $m_{S_1^0} \lesssim 30-40 $ TeV.

\subsection{Direct detection}
\label{sec:directdetectionnum}

\begin{figure}[!tb]
\centering
    \includegraphics[width=0.48\textwidth]{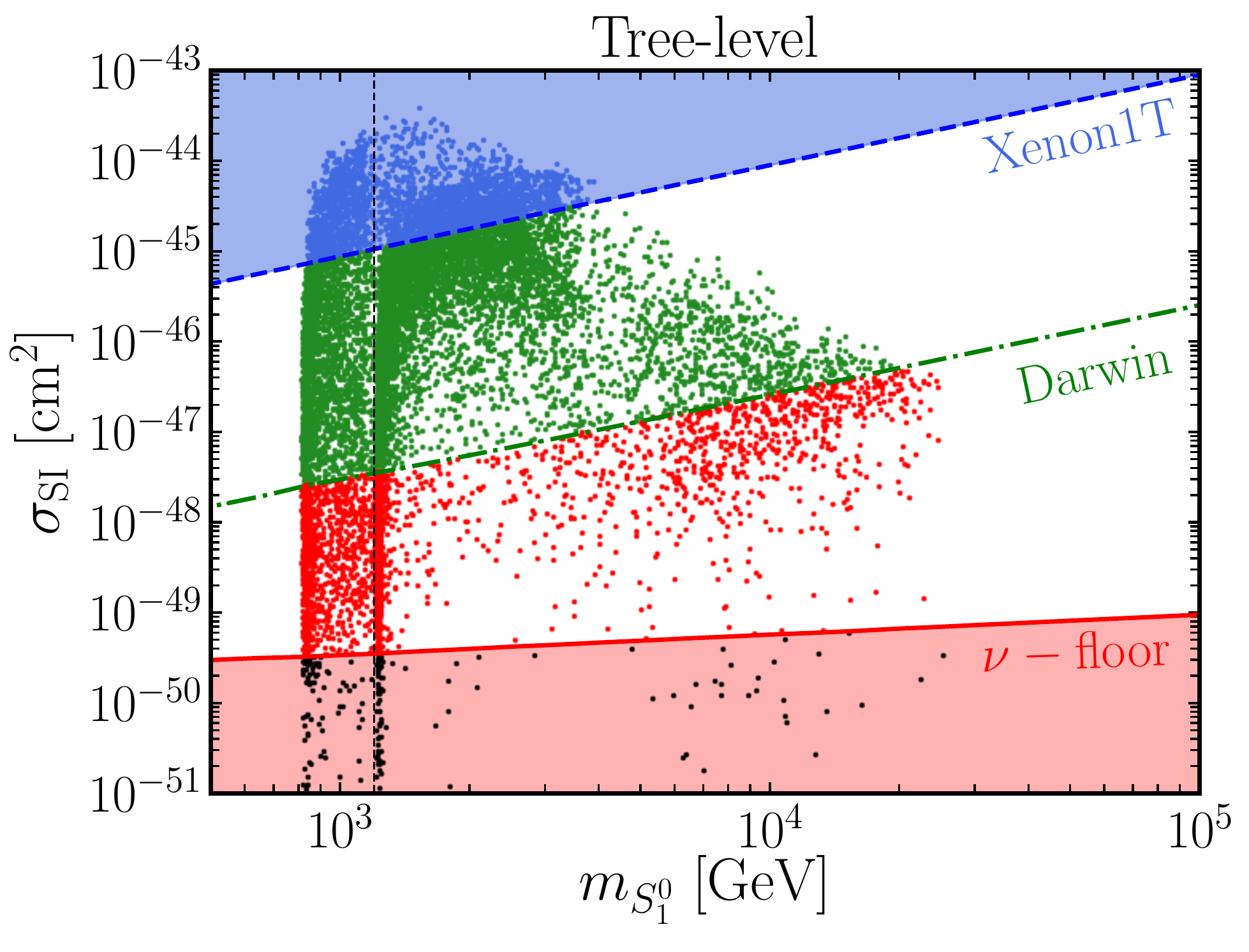}
    \includegraphics[width=0.48\textwidth]{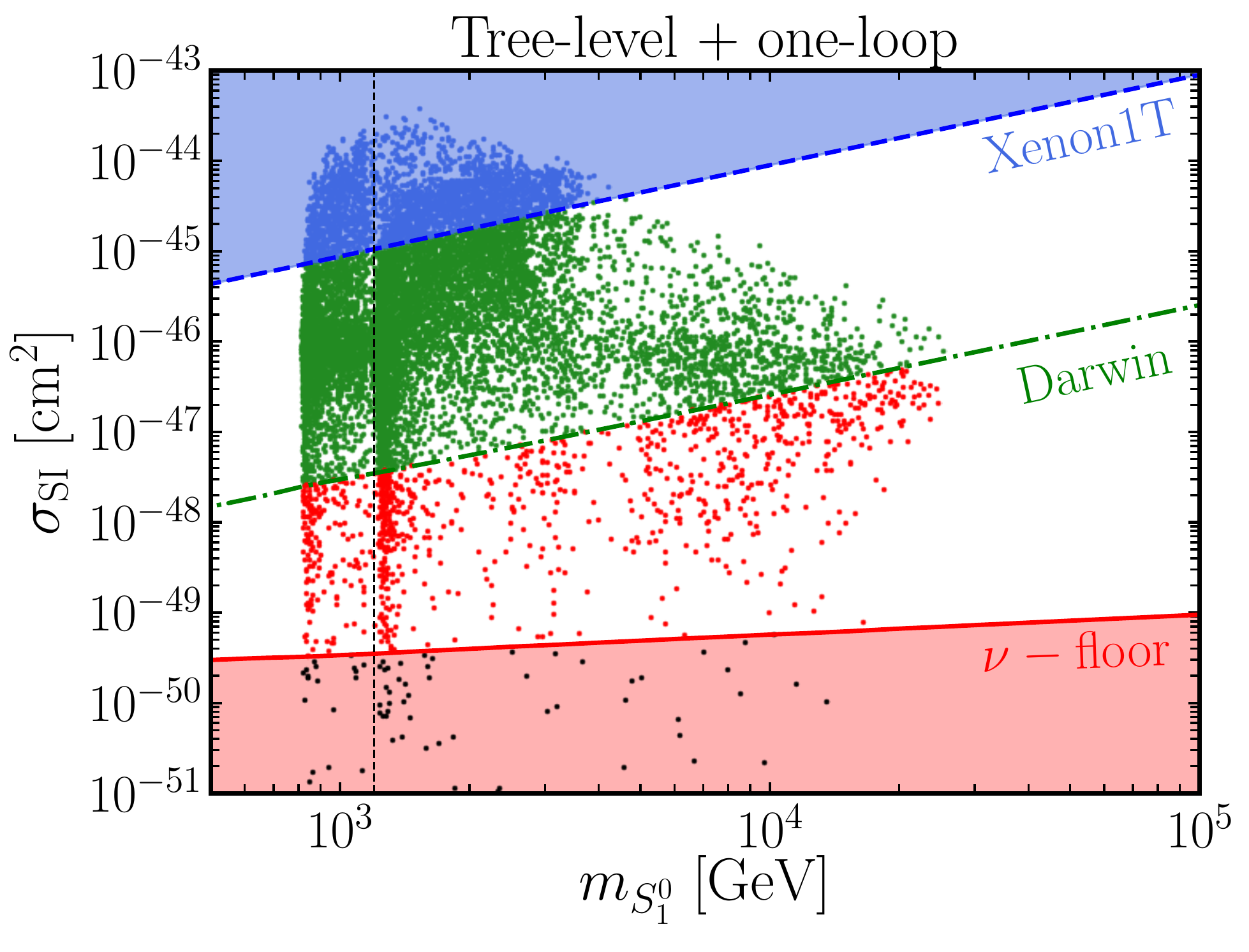}
    \caption{Dark matter spin independent direct detection cross section as a function of the lightest neutral scalar mass. Constraints from the Xenon1T experiment~\cite{Aprile:2018dbl} are shown in blue in addition to the future sensitivity achievement of the upcoming Darwin experiment~\cite{Aalbers:2016jon} in green and the so-called neutrino floor~\cite{Billard:2013qya} in red. Blue dots are excluded by Xenon1T. Green dots will be probed by Darwin. Red (black) dots correspond to points above (below) the neutrino floor.}
    \label{fig:DD}
\end{figure}

Values of the DM spin independent nucleon cross section, predicted by our scan in the parameter space, are depicted in Fig.~\ref{fig:DD} as a function of the mass of the DM candidate $m_{S_1^0}$. In addition, we represented constraints from the Xenon1T experiment~\cite{Aprile:2018dbl} in blue in addition to the future sensitivity achievement for the upcoming Darwin experiment~\cite{Aalbers:2016jon} in green and the so-called neutrino floor in red~\cite{Billard:2013qya}. 

The left panel of Fig.~\ref{fig:DD} shows the tree-level contribution from Eq.~(\ref{eq:sigmaDD_tree}). The triangle-like shape of the cluster of points is roughly delimited by a vertical line around $ m_{S_1^0} \simeq 750 $ GeV, and two diagonal line converging at around $m_{S_1^0} \simeq 30$ TeV, corresponding to three different limiting effects. For small masses $m_{S_1^0}\lesssim 750$ GeV, co-annihilations are too efficient to yield the correct relic abundance. The upper diagonal limit is set by the perturbative limit reached by dimensionless couplings of the scalar potential and the Yukawa couplings as discussed in the previous subsection. The requirement of reproducing the neutrino masses and mixings as described in Sec.~\ref{sec:numasses} imposes a lower bound on the coupling $\kappa $ around $\kappa \gtrsim 10^{-3}$ for $m_{S_1^0} \sim 1$ TeV and $\kappa \gtrsim 10^{-1}$ for $m_{S_1^0} \sim 10$ TeV which translates into the lower diagonal limit in Fig.~\ref{fig:DD}. Almost the entire part of the parameter space lies in this triangular contour, apart from very few funnel points.

The right panel of Fig.~\ref{fig:DD} shows the sum of the tree-level and one-loop electroweak contributions from Eq.~(\ref{eq:sigmaDD_tot}). The loop contribution generate values for the SI cross section of the order of $10^{-47}-10^{-46}~\text{cm}^2$ which can be seen as a more clustered region around these values. However, interferences between the tree-level and electroweak contributions tend to scatter the point to lower values of the cross section while leaving unmodified points corresponding to cross section larger than $10^{-47}-10^{-46}~\text{cm}^2$ at tree-level.

Interestingly, a sizable part of the parameter space is already excluded by the Xenon1T experiment. In addition, almost the entire parameter space lies above the neutrino-floor including a majority of points that should be probed by the Darwin experiment, implying that the model could be almost entirely tested within the next decades.  \par\medskip

\subsection{Indirect detection}

\begin{figure}[!t]
\centering
    \includegraphics[width=0.75\textwidth]{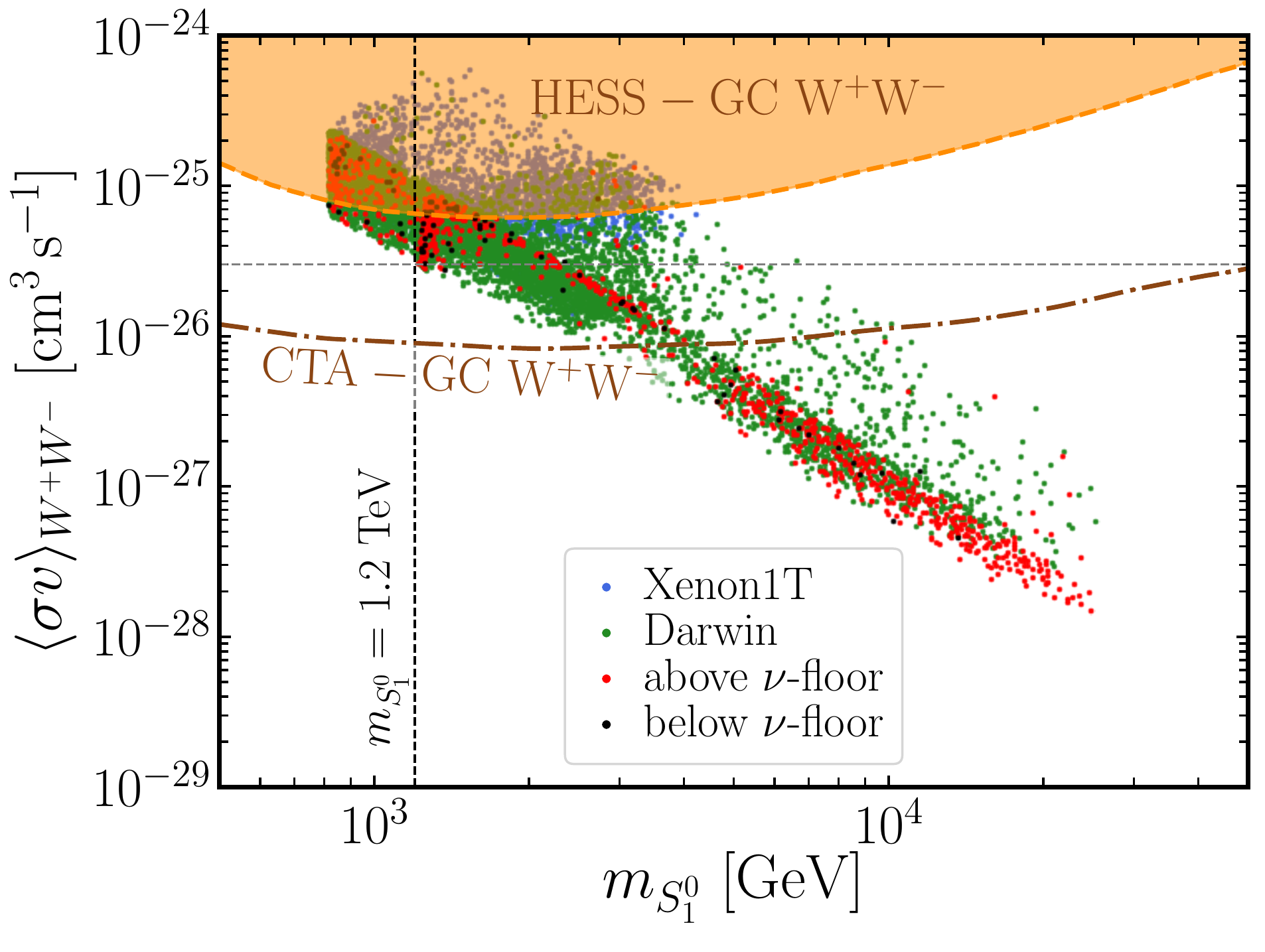}
    \caption{Dark matter perturbative velocity averaged annihilation cross section into $W^+ W^-$ and constraints from HESS~\cite{Abdallah:2016ygi} as well as sensitivity prediction for CTA~\cite{CTAConsortium:2018tzg}. The various points represent the results from our scan in the parameter space reproducing both the correct dark matter relic density and neutrino masses.
    The color code is the same as in Fig.~\ref{fig:DD}, as described in Sec.~\ref{sec:DD}.}
    \label{fig:IDWW}
\end{figure}

We represented in Fig.~\ref{fig:IDWW} the points from our scan in the parameter space, allowing to reproduce the observed DM relic abundance and neutrino masses, projected in the plane $( m_{S_1^0},\langle \sigma v \rangle)$ as well as the constraint from HESS~\cite{Rinchiuso:2017pcx}. Values of $\langle \sigma v \rangle$ are computed numerically using \texttt{micrOMEGAS}~\cite{Belanger:2001fz,Belanger:2018ccd}. In addition we depicted the CTA sensitivity projection~\cite{CTAConsortium:2018tzg}, assuming 500h of observations towards the Galactic Center and DM annihilating to $W^+W^-$ final state. Similar analyses have reached the same conclusion~\cite{Pierre:2014tra,Silverwood:2014yza,Lefranc:2015pza}, i.e. for DM annihilations into $\bar b b$, $\tau^+ \tau^-$ and  $W^+W^-$, CTA should probe the vanilla value of the velocity averaged annihilation cross section $\langle \sigma v \rangle = 3 \times 10^{-26}$. We used the same color code than for Fig.~\ref{fig:DD}, as described in Sec.~\ref{sec:DD}. The cluster of points in Fig.~\ref{fig:IDWW} is roughly shaped as a ``broad line" decreasing towards high DM masses. This line follows essentially the behavior of the $W^+W^-$ cross section of Eq.~(\ref{eq:sigmavWW}) as a function of the DM mass. The broadness of this line is determined by coannihilations that allow for several values for $\langle \sigma v \rangle_{W^+ W^-}$, for a given DM mass while still satisfying the relic density condition. The spread of points in particular for DM masses $1<m_{S_1^0}<4$ TeV are to additional scalar-potential couplings contributing to $\langle \sigma v \rangle_{W^+ W^-}$ such as $\kappa$ or $\lambda_{H \Omega}$, as can be deduced from Eq.~(\ref{eq:sigmavWW}), being relatively large as for this part of the parameter space, as discussed previously. \par \medskip
Fig.~\ref{fig:IDWW} shows that HESS is already constraining a sizable part of the parameter space which should be substantially improved by CTA. 
It is worth notice here that a large proportion of the points lying below the neutrino-floor in Fig.~\ref{fig:DD} seems to correspond to a relatively high value for $\langle \sigma v \rangle_{W^+ W^-}$ that are already excluded by HESS or should be in the near future by CTA. As those points cannot be probed with the future generation of xenon-based direct detection experiments, indirect searches offer a very interesting complementary discovery prospects for this model. \par \medskip

\begin{figure}[!t]
\centering
    \includegraphics[width=0.48\textwidth]{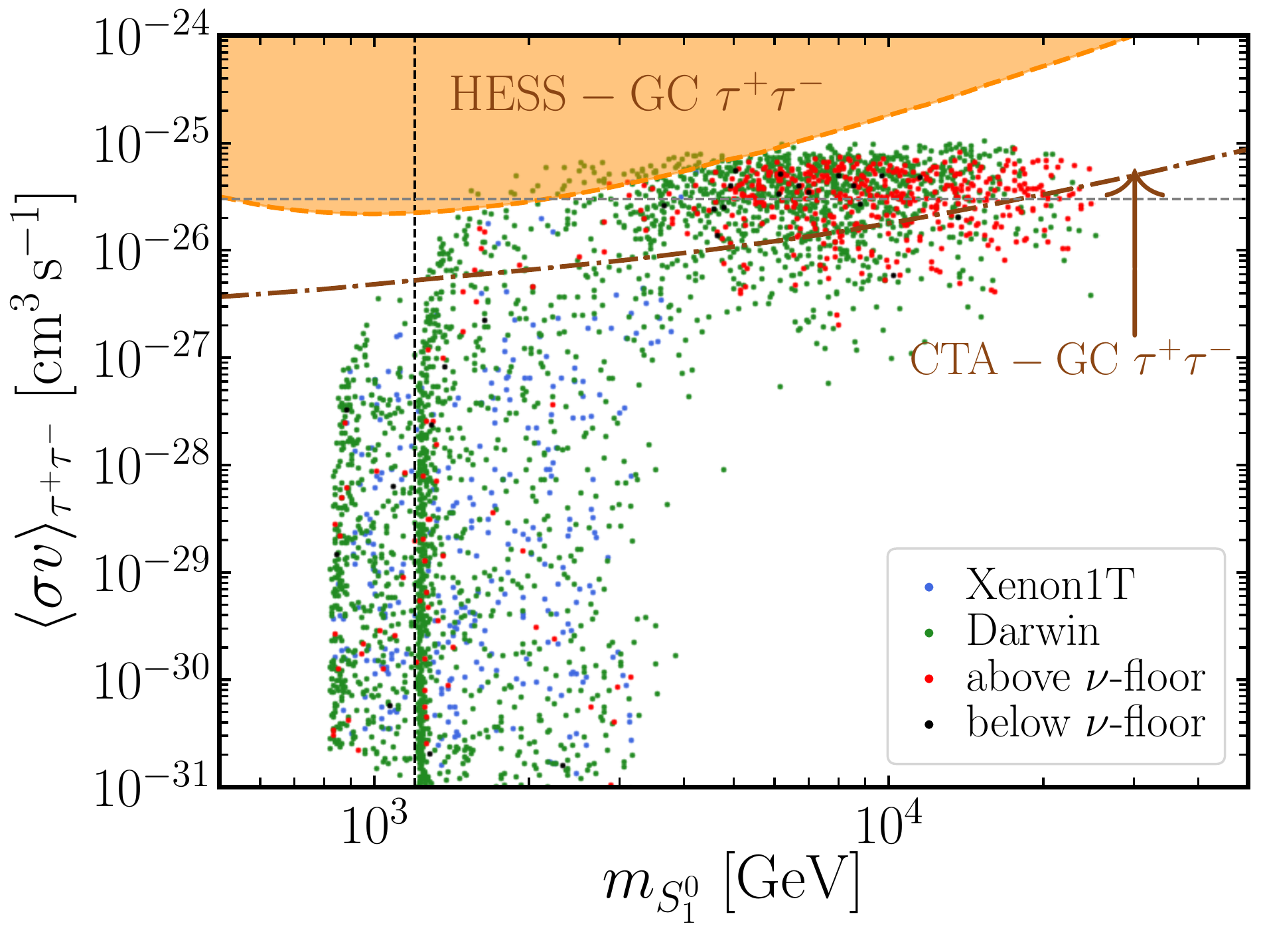}
    \includegraphics[width=0.48\textwidth]{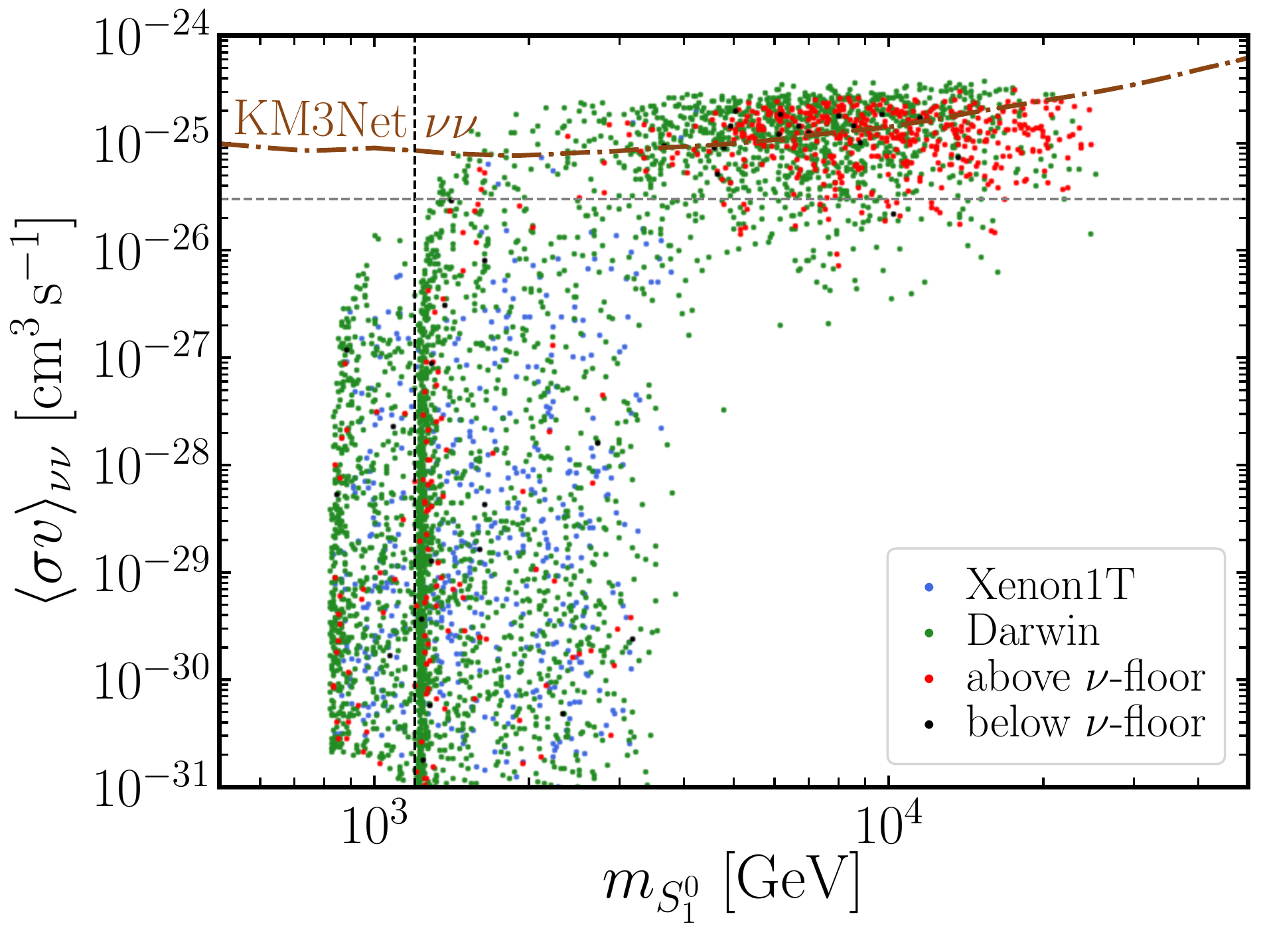}
    \caption{Dark matter perturbative velocity-averaged annihilation cross section into $\tau^+ \tau^-$ and $\nu \nu$. On the left, limits from HESS~\cite{Rinchiuso:2017pcx} are represented by a dashed orange line. The CTA sensitivity estimate of~\cite{Carr:2015hta}, assuming 500h of observation towards the galactic center, is depicted in dash-dotted line. On the right, sensitivity prospects for KM3Net~\cite{2019ICRC...36..552G,Arguelles:2019ouk} are shown in dash-dotted line. The color code is the same as in Fig.~\ref{fig:DD}, as  described in Sec.~\ref{sec:DD}.}
    \label{fig:IDleptons}
\end{figure}

In addition, as commented in Sec.~\ref{sec:directdetectionnum}, DM annihilations to leptons such as $\tau^+ \tau^-$ and $\nu \nu$ are significant for large DM masses $m_{S_1^0} \gtrsim 3-4$ TeV as the gauge annihilation channel is suppressed, as can be seen in Fig.~\ref{fig:IDWW}. We represented the annihilation cross section to $\tau^+ \tau^-$ and $\nu \nu$ in Fig.~\ref{fig:IDleptons} using the same color code than Fig.~\ref{fig:DD}. In addition, we represented as well the CTA sensitivity estimate of~\cite{Carr:2015hta}, assuming 500h of observation towards the galactic center on the left panel and the sensitivity prospects for KM3Net~\cite{2019ICRC...36..552G,Arguelles:2019ouk} on the right panel. Interestingly the points located below the Darwin sensitivity projection or below the neutrino floor, at large DM masses, tends to be accompanied by sizable leptonic annihilation cross section that should be probed by CTA and KM3Net. 

Long range forces mediated by electroweak gauge bosons between DM states with masses $m_{S_1^0} \gg m_{W}, m_{Z}$  can alter the perturbative prediction for the DM annihilation cross section, the well-know effect known as Sommerfeld enhancement~\cite{Hisano:2003ec,Hisano:2006nn}. Formation of bound states, in addition to introducing new annihilation channels~\cite{Asadi:2016ybp}, can also manifest in resonances in the Sommerfeld enhancement, in particular for small velocities $\langle v_{\rm DM} \rangle \ll 1$. It has been shown in Ref.~\cite{Hisano:2006nn} that the Sommerfeld enhancement is not numerically significant at the time of freeze-out, typically occuring at $m_{S_1^0}/T \simeq 20$, therefore should not affect strongly our prediction for the relic density. However this effect could increase the cross section relevant for indirect detection by a factor of $10^2-10^3$~\cite{Hisano:2006nn,Cirelli:2007xd,Blum:2016nrz} for a typical pure-triplet DM candidate with a mass above the TeV scale and away from resonances, but could be even larger for specific masses. This effect should be taken into account to accurately assess the viability of the model. However, given the complexity of the model and the fact that our DM candidate is a mixed state, a dedicated analysis is required to numerically estimate the Sommerfeld enhancement in a reliable way, which is beyond the scope of this paper. 
\par \medskip

As a result, the complementarity of future sensitivity achievement on the spin independent from Darwin and annihilations within the galactic halo to leptons by CTA and KM3Net offers a very interesting discovery prospects from this model addressing simultaneously the problem of the dark matter abundance and neutrino masses. In addition to searches with CTA, the future Southern Wide-field Gamma-ray Observatory (SWGO)~\cite{Abreu:2019ahw} in the sourthern hemisphere could have a better sensitivity to the galactic center and other regions of the gamma-ray sky~\cite{Viana:2019ucn}. Taking into account the Sommerfeld enhancement would essentially boost the indirect detection signal and reinforce the optimistic discovery prospect message that we are addressing here.

\section{Additional constraints}
\label{sec:addconstraints}

\paragraph{Radiative breaking of the $\mathbb{Z}_2$ symmetry}
Large fermionic masses could be responsible for breaking the discrete $\mathbb{Z}_2$ symmetry by loop effects. Indeed, as shown in~\cite{Merle:2016scw}, the $\beta$-functions of the mass parameters of the scalar potential in the original scotogenic model could receive negative contributions from the fermionic states and drive the mass parameters towards negative values at some high energy scale, resulting in a breaking of the $\mathbb{Z}_2$ symmetry. Depending on the $\mathbb{Z}_2$ breaking scale, consistency of the low energy theory could be spoiled and affect the DM stability and density production. As shown in~\cite{Merle:2015gea}, additional $\mathbb{Z}_2$-odd scalars could help to stabilize the behavior of the $\beta$ functions. The complete numerical RGE analysis lies beyond the scope of this paper. However, as in our case the DM candidate is the lightest $\mathbb{Z}_2$-odd state, such effects should occur at scales higher than the DM freeze-out temperature and would not affect the DM abundance. \par \medskip

\paragraph{Neutrinoless double beta decay} Assuming normal ordering and the lightest neutrino massless, as described in Sec.~\ref{sec:numasses}, then there is only one physical Majorana phase $\phi$ in the neutrino sector. In this cases, the effective mass parameter characterizing the amplitude for neutrinoless double beta decay can be expressed as~\cite{Rodejohann:2011vc,Avila:2019hhv,Zyla:2020zbs}

\begin{equation}
    m_{\beta \beta}  \,=\,\bigg\rvert \sum_{i=1,2,3} m_{\nu_i} U_{ie}^2 \bigg\rvert \,=\, \left| |m_{\nu_2}| \sin ^2(\theta_{12}) \cos ^2(\theta_{13}) + |m_{\nu_3}|  \sin ^2(\theta_{13}) e^{-2 i \phi } \right| \,,
\end{equation}
where $\phi$ is the only free parameter after imposing the model to reproduce neutrino parameters described in Sec.~\ref{sec:numasses}. Numerical values for this parameter are within the range $  m_{\beta \beta}   \in [0.0015,0.0038] \,\,\text{eV}$ which is below the sensitivity of the KamLAND-Zen experiment~\cite{KamLAND-Zen:2016pfg} and below the sensitivity that could be achieved by the nEXO experiment in the future $m_{\beta \beta} \lesssim 6$ meV~\cite{Albert:2017hjq}.\par \medskip

\paragraph{Oblique parameters} 
The new scalar and fermionic multiplet contribute to the vacuum polarization of electroweak gauge bosons and shift their SM expected values, which translates into a shift of the oblique parameters. In particular in scotogenic-like models, deviations from the $T-$parameter can be sizable for large mass scales and mass splittings. Assuming vanishing mixing angles and small relative mass splitting between the charged and neutral components $\delta \equiv (m_{S^{+(+)}}-m_{S^0})/m_{S^0}\ll 1$, the contribution from the $Y=1$ scalar triplet to the $T-$parameter is of order~\cite{Lavoura:1993nq,Chen:2019okl}
\begin{equation}
    T\, \simeq \, \frac{ \delta^2  m_{S^0}^2}{6 \pi  s_W^2 c_W^2  m_Z^2} \, \simeq \, 0.075 \, \left( \dfrac{\delta}{0.05} \right)^2\left( \dfrac{m_{S^0}^2}{1 \,\text{TeV}} \right)^2\,,
\end{equation}
which remains below the current experimental uncertainty $T=0.03\pm0.12$~\cite{Zyla:2020zbs} for our viable parameter space. Numerical values of the mass splitting can be found in Appendix~\ref{sec:additional_plots}. In addition, as detailed in Appendix~\ref{sec:fermion_mass_splitting}, the mass splitting for the neutral and charged components of the vector-like lepton doublet is typically small $\Delta m_f\sim 10^{-4} m_f$ for $m_f=1$ TeV, therefore does not contribute significantly to the $T-$parameter.\par \medskip

\paragraph{Collider constraints}

Charged scalars can affect the effective coupling involved in the $h\rightarrow \gamma \gamma$ process~\cite{Arhrib:2011vc} which is constrained to be around the SM-expected value. Since the viable part of our parameter space features many scalars with similar masses typically between $\sim 800$ GeV and $\sim 30$ TeV, this model offers interesting complementary signatures at high luminosity colliders. For instance in proton-proton collisions, a DM pair could be produced in association with two or four charged leptons via $pp\rightarrow \gamma^*, Z^* \rightarrow \bar f^{\pm *} f^{\pm *} \rightarrow \ell^+ \ell^- S^0_1 S^0_1$ or $pp\rightarrow \gamma^*, Z^* \rightarrow  S^{\pm \pm *} S^{\pm \pm *} \rightarrow \ell^+\ell^+ \ell^- \ell^- S^0_1 S^0_1$ where the DM pairs would be interpreted as missing transverse energy at collider. However, notice here that the typical $S^{++} \rightarrow \ell^+\ell^+$ signal expected from models extending the TII-seesaw~\cite{Farzan:2010mr} should not be present in our case as the $\mathbb{Z}_2$ symmetry is unbroken and forbid processes with a even number of $\mathbb{Z}_2$-odd states. On the other hand, the following processes are present in our model $S^{++} \rightarrow \ell^+\ell^+ S^0_i $ and $S^{++} \rightarrow f^+\ell^+ $  which somehow could be interesting to analyze. 
Analyses of such processes have been performed in similar models~\cite{Beniwal:2020hjc,Jana:2020joi,Chakraborti:2020zxt,Hagedorn:2018spx,Avila:2019hhv}. However, their results cannot be applied directly to our case as the quantum numbers of the particles present in the theory are different. Such processes in our model would require a dedicated investigation which goes beyond the scope of this paper and is left for future work.

\section{Conclusion}
\label{sec:conclusions}

In this work, we investigated the phenomenology of a model that provides a Dark Matter candidate and a mechanism to generate neutrino masses using as inspiration the fields in the Type-II seesaw. In this model, we introduced, in addition to the SM fields, a pair of $SU(2)_L$ scalar triplets with hypercharge $Y=1$ and $Y=0$ and a $SU(2)_L$ doublet vector-like fermion. All the extra fields are considered as charged under a  discrete $\mathbb{Z}_2$ symmetry. Three of these particles (two scalars $S^0_1$, $\widetilde{S}^0$ and one fermion $f^0$) are electrically neutral and depending on the specific parameters, could be the lightest state charged under the $\mathbb{Z}_2$ symmetry, i.e. a  WIMP-like dark matter candidate. 

Various phenomenological aspects of this model were investigated. We showed that light neutrino masses, generated by loop contributions from $\mathbb{Z}_2-$odd states, could be accommodated in this model, in addition to the flavour structure reproducing the observed neutrino oscillations patterns. In this setup, two neutrinos acquire non-zero masses while one remains massless. We focused on reproducing the normal hierarchy for neutrino masses but inverted hierarchy is not excluded by our analysis and could still be viable in this framework.

We identified the points in the parameter space allowing to reproduce the dark matter relic abundance as observed by the Planck collaboration. In addition to studying the theoretical consistency of the model, we explored dark matter direct detection and indirect searches signatures. In addition, we considered constraints from the lepton flavour violating process $\mu \rightarrow e \gamma$. Part of the parameter space, satisfying the correct relic density condition and $\mu \rightarrow e \gamma$ constraints, is in tension with current bounds from gamma-ray observations from HESS ($W^+ W^-$ and $\tau^+ \tau-$ annihilation channels) and DM-nuclei scattering bounds from the Xenon1T experiment. We showed that among the three possible dark matter candidates, only one of them, the CP-even scalar $S^0_1$, satisfies these constraints. The phenomenological viable mass for this DM candidate ranges from 700~GeV to 30~TeV. We identified points in the parameter space within the reach of the next generation of direct detection and indirect detection experiments. In particular, we showed that a large proportion of the parameter space should be probed by the Darwin experiment while a sizable part should evade these bounds but would still remain above the neutrino floor, i.e. accessible with xenon-based experiments in the future. Nevertheless, a minority of these points lie below the neutrino floor.  Interestingly, an important proportion of these points lying below the neutrino floor would give rise to $W^+W^-$ annihilation signals that should be observed by CTA and perhaps the Southern Wide-field Gamma-ray Observatory in the future. 

In addition, we showed the complementarity between direct detection and indirect signals from DM annihilations to $\tau^+ \tau^-$ by CTA and to $\nu \nu$ by KM3Net. Such various and complementary signatures would allow to verify or refute the model in the future, but also to discriminate between this model and others, in case some signal is observed. The Sommerfeld enhancement was not considered in this model but is necessary to predict more accurately the expected indirect detection signal. Nevertheless, It would only reinforce the optimistic detection prospect that we are addressing here and this computation is left for future work.

Moreover, the model presents features that would give rise to additional interesting signatures at present and future colliders such as new charged scalars and vector-like fermions. In addition, the new states considered in this setup could impact additional lepton-flavour-violating observables and offer a complementary way of probing the model considered in this work. The detailed analysis of these signatures is left for future work. \par

%\section*{Acknowledgments}
\acknowledgments
The authors would like to thank Nicolas Rojas for his implication in the early stages of this project. The authors would like to thank Viviana Gammaldi for useful discussions, Renato Fonseca for discussions about BFB conditions, and Avelino Vicente for useful discussions about lepton-flavour-violation. The work of MP was supported by the Spanish Agencia Estatal de Investigaci\'{o}n through the grants FPA2015-65929-P (MINECO/FEDER, UE),  PGC2018-095161-B-I00, IFT Centro de Excelencia Severo Ochoa SEV-2016-0597, and Red Consolider MultiDark FPA2017-90566-REDC. MP would also like to thank the Paris-Saclay Particle Symposium 2019 with the support of the P2I and SPU research departments and the P2IO Laboratory of Excellence (program "Investissements d'avenir" ANR-11-IDEX-0003-01 Paris-Saclay and ANR-10-LABX-0038), as well as the IPhT for hospitality during part of the realization of this work.
RL thanks the Instituto de Física Teórica (UAM/CSIC) for the hospitality. RL was supported by Universidad Católica del Norte through the Publication Incentive program No. CPIP20180343 and CPIP20200063.

%%%%%%%%%%%%%%%%%%%%%%%%%%%%%%%%
%%%%%%%% APPENDICES
%%%%%%%%%%%%%%%%%%%%%%%%%%%%%%%%
\appendix

\section{Rotation matrices}
\label{sec:rotation}
In this appendix we present the parameterization used for the diagonalization of the CP-even neutral scalar and charged scalars.

\subsection{CP-even neutral scalars}
In the gauge basis $(\Omega^0_R,\Delta^0_R)$, the mass matrix is diagonalized through a basis rotation in such a way that
\begin{equation}
    R_0^T \mathcal{M}^2_0 R_0 = \text{diag}\big(m^2_{S^0_1}, m^2_{S^0_2}\big) \, ,
\end{equation}
where the mass matrix eigenvalues are ordered following $m^2_{S^0_1} \le m^2_{S^0_2}$, and the rotation matrix $R_0$ satisfies the condition $R_0^T R_0 = R_0 R_0^T = \mathbbm{1}$. The relation between the mass states and the gauge basis is simply:
\begin{equation}
\left(
\begin{array}{c}
 \Omega_R^0 \\
 \Delta^0_R \\
\end{array}
\right)= R_0 \left(
\begin{array}{c}
 S^0_1 \\
 S^0_2 \\
\end{array}
\right) \, ,
\end{equation}
where the most common parameterization for rotation matrix is in base on trigonometric functions 
\begin{equation}
R_0 = \left(
\begin{array}{cc}
 \cos \theta_0 &  \sin \theta_0 \\
  -\sin \theta_0 &  \cos \theta_0 \\
\end{array}
\right) \, .
\end{equation}
For the case of the CP-even neutral scalar, the mixing angle $\theta_0$ is obtained from the relation:
\begin{equation}
    \tan \big( 2\theta_0 \big) \, = \, \dfrac{  s_\kappa \kappa  v_h^2}{2(m_\Delta^2-m_\Omega^2)+v_h^2 \left(\lambda_{H\Delta} -\lambda_{H\Omega} + \lambda_{H\Delta}^\prime \right)}\,.
\end{equation}
However the mixing angle in the latter expression could be shifted in $\pi$ to ensure on the ordering of the mass matrix eigenvalues. Using this parameterization, the limit when $\theta_0\mod{(\pi)} \rightarrow 0$, implies that $S^0_1 \rightarrow \Omega^0_R$  and $S^0_2 \rightarrow \Delta^0_R$; and the opposite case occurs when $\theta_0\mod{(\pi)} \rightarrow \pi/2$ implying $S^0_1 \rightarrow \Delta^0_R$  and $S^0_2 \rightarrow \Omega^0_R$.

Complementary, another parameterization for the rotation matrix $R_0$ can produce sorted eigenvalues if we assume:
\begin{equation}
R_0 = \left(
\begin{array}{cc} 
 \sqrt{\dfrac{1 + \psi_0}{2}} &  - s_\kappa \sqrt{\dfrac{1 - \psi_0}{2}} \\
   s_\kappa \sqrt{\dfrac{1 - \psi_0}{2}} &  \sqrt{\dfrac{1 + \psi_0}{2}} \\
\end{array}
\right) \, ,
\end{equation}
and if the CP-even neutral scalar mass matrix in the gauge basis Eq.~(\ref{eq:neutral_massmatrix}) is written as follows:
\begin{equation}
\mathcal{M}_0^2 = \left(
\begin{array}{cc}
 A_0 &  s_\kappa B_0 \\
   s_\kappa B_0 &  C_0 \\
\end{array}
\right) \, ,
\end{equation}
where $s_\kappa = \pm 1$ is the sign of $\kappa$ present in the mass matrix. Notices that $A_0$, $B_0$, and $C_0$ are positive. 
Under these assumptions, we get that
\begin{equation}
    \psi_0 = \frac{C_0 - A_0}{\sqrt{4 B_0^2 + (C_0 - A_0)^2}} \, ,
\end{equation}
always produces sorted eigenvalues:
\begin{eqnarray}
    m^2_{S^0_1} &=& \frac{1}{2} \left( C_0 + A_0 - \sqrt{4 B_0^2 + (C_0 - A_0)^2} \right) \, ,\\
    m^2_{S^0_2} &=& \frac{1}{2} \left( C_0 + A_0 + \sqrt{4 B_0^2 + (C_0 - A_0)^2} \right) \, .
\end{eqnarray}
Using this parameterization, $\psi_0 = 1$ implies the alignment $S^0_1 \rightarrow \Omega^0_R$  and $S^0_2 \rightarrow \Delta^0_R$ and $\psi_0 = -1$ implies $S^0_1 \rightarrow \Delta^0_R$  and $S^0_2 \rightarrow \Omega^0_R$. The connection between both parameterizations is straightforward:
\begin{eqnarray}
    \psi_0 &=& \cos(2 \theta_0) \, ,\\
    \sin(\theta_0) &=&  -s_\kappa \sqrt{\dfrac{1 - \psi_0}{2}} \, .
\end{eqnarray}

%%%%%%%%%%%%%%%%%%%%%%%%%%%%%%%%%%%%%%%%%%%%%%%%%%%%%%%%%%%%%%%%%%%%%%%%%%%%%%%%%%%
\subsection{Charged scalars with \texorpdfstring{$Q=1$}{Q=1}}
Similarly to the neutral scalar, the mass matrix in the gauge basis $(\Omega^\pm,\Delta^\pm)$ is diagonalized by a rotation:
\begin{equation}
    R_1^T \mathcal{M}^2_\pm R_1 = \text{diag}\big(m^2_{S^\pm_1}, m^2_{S^\pm_2}\big) \, ,
\end{equation}
where the transformation between the gauge basis and the mass basis is:
\begin{equation}
    \left(
\begin{array}{c}
 \Omega^\pm \\
 \Delta^\pm \\
\end{array}
\right)= R_1 \left(
\begin{array}{c}
 S^\pm_1 \\
 S^\pm_2 \\
\end{array}
\right) \, .
\end{equation}
The $R_1$ parameterization in terms of trigonometric functions is 
\begin{equation}
R_1 = \left(
\begin{array}{cc}
 \cos \theta_1 &  \sin \theta_1 \\
  -\sin \theta_1 &  \cos \theta_1 \\
\end{array}
\right) \,  ,
\end{equation}
where the mixing angle can be expressed as:
\begin{equation}
    \tan \big( 2\theta_1\big) \, = \, \dfrac{- s_\kappa \sqrt{2}\kappa  v_h^2}{ 4(m_\Delta^2- m_\Omega^2)+v_h^2 \left(\lambda_{H\Delta}-2 \lambda_{H\Omega} +2 \lambda_{H\Delta}^\prime \right)}\, .
\end{equation}
Notice that the mixing angle $\theta_1$ might have a $\pi$ shift due to the ordering of the eigenvalues.

Complementary to the previous parameterization, the mass matrix (Eq.~\ref{eq:charged_massmatrix}) can be written as:
\begin{equation}
    \mathcal{M}^2_\pm = \left(
        \begin{array}{cc}
        A_1 &  -s_\kappa B_1 \\
        -s_\kappa B_1 &  C_1 \\
        \end{array}
\right) \, ,
\end{equation}
where $A_1$, $B_1$, and $C_1$ are positive; and $s_\kappa$ is the sign of $\kappa$. The corresponding rotation matrix must be:
\begin{equation}
R_1 = \left(
\begin{array}{cc} 
 \sqrt{\dfrac{1 + \psi_1}{2}} &   s_\kappa \sqrt{\dfrac{1 - \psi_1}{2}} \\
   - s_\kappa \sqrt{\dfrac{1 - \psi_1}{2}} &  \sqrt{\dfrac{1 + \psi_1}{2}} \\
\end{array}
\right) \, .
\end{equation}
Notice the difference by a relative sign with respect to the neutral-scalar case. Here, it is easy to see that:
\begin{equation}
    \psi_1 = \frac{C_1 - A_1}{\sqrt{4 B_1^2 + (C_1 - A_1)^2}} \, ,
\end{equation}
produces ordered eigenvalues such that:
\begin{eqnarray}
    m^2_{S^\pm_1} &=& \frac{1}{2} \left( C_1 + A_1 - \sqrt{4 B_1^2 + (C_1 - A_1)^2} \right) \, ,\\
    m^2_{S^\pm_2} &=& \frac{1}{2} \left( C_1 + A_1 + \sqrt{4 B_1^2 + (C_1 - A_1)^2} \right) \, .
\end{eqnarray}
The connection between both parameterizations is straightforward:
\begin{eqnarray}
    \psi_1 &=& \cos(2 \theta_1) \, ,\\
    \sin(\theta_1) &=&  s_\kappa \sqrt{\dfrac{1 - \psi_1}{2}} \, .
\end{eqnarray}

%%%%%%%%%%%%%%%%%%%%%%%%%%%%%%%%%%%%%%%%%%%%%%%%%%%%%%%%%%%%%%%%%%%%%%%%%%%%%%%%%%
\section{Scan strategy of the parameter space}
\label{sec:scan_strategy}
In this appendix we detail how the various relevant parameters of the Lagrangian can be expressed in terms of a set of convenient variables introduced to perform efficiently a scan over the parameter space. 

\subsection{Scalar sector}
Using parametrization of the ration in terms of $\psi_{0,1}$ defined in the previous section, we define the convenient dimensionless quantities
\begin{equation}
    \phi_{0,1}=\sqrt{1 - \psi_{0,1}^2}\,,
\end{equation}
and
\begin{align}
     \alpha_0 &\,\equiv\,  \dfrac{v_h^2 \kappa}{2 m_{S_1^0}^2 \phi_0} \, , \quad     \alpha_1 \,\equiv\, \dfrac{v_h^2 \kappa}{2 m_{S_1^0}^2(1+\alpha_{01})\phi_2}\, , \\ 
     \alpha_{01} &\,\equiv\,  \dfrac{v_h^2 \kappa}{8\phi_1 \phi_0 m_{S_1^0}^2} \left(2\phi_1(1-\psi_0) -\sqrt{2}\phi_0(1-\psi_1) \right) \, ,
\end{align}
which allows to express the coupling
\begin{eqnarray}
    \lambda_{H \Delta} \,= \,\kappa\left(2\dfrac{\psi_0}{\phi_0} - \sqrt{2} \dfrac{\psi_1}{\phi_1} \right) \, ,
\end{eqnarray}
and relates neutral scalar mass eigenvalues by the relations
\begin{equation}
    m_{S_2^0} \,=\,  m_{S_1^0} \sqrt{1 + \alpha_0}\, , \quad \quad \quad
    m_{S_1^\pm} \,=\,   m_{S_1^0}\sqrt{1 + \alpha_{01}}\, ,
\end{equation}
and
\begin{align}
    m_{\tilde{S}^0} &\,=\,  m_{S_1^0} \sqrt{1 + \frac{\alpha_0}{2}(1 + \psi_0)}\, , \quad
    m_{S_2^\pm} \,=\,  m_{S_1^\pm}  \sqrt{1 + \frac{\alpha_1}{\sqrt{2}}} \, , \\ 
    m_{S^{\pm \pm}} &\,=\,  \sqrt{m_{\tilde{S}^0}^2 - \frac{1}{2}\lambda_{H \Delta}v_h^2 }  \, .
    \label{eq:mass_scalars_scan_parametrization}
\end{align}
Mass parameters of the scalar potential can be expressed in terms of these parameters as
\begin{align}
    m_\Omega^2  \,=\, & \dfrac{1}{2}\left(m_{S_1^0}^2\big(2 + \alpha_0 (1 - \psi_0)\big) - v_h^2 \lambda_{H \Omega}\right) \, , \nonumber \\
    m_\Delta^2  \,=\, & \dfrac{1}{2}\left(m_{S_1^0}^2\big(2 + \alpha_0 (1 + \psi_0)\big) - v_h^2 (\lambda_{H \Delta}+\lambda_{H \Delta}^\prime)\right) \, .
\end{align}
As a result, the parameters of the scalar potential of Eq.~(\ref{eq:scalarpotential}) can be expressed in terms of the set of variables $\{\kappa, s_\kappa, \lambda_{\Omega}, \lambda_{\Delta}, \lambda_{\Delta}^\prime,  \lambda_{H \Omega}, \lambda_{H \Delta}^\prime, \lambda_{ \Delta \Omega}, \psi_0, \psi_1\}$.

\subsection{Neutrino sector} \label{sec:appendix_neutrino_masses}
In order to reproduce neutrino masses and mixing angles as described in Sec.~\ref{sec:numasses}, we define the quantity  $\phi_N \equiv \arctan \left[\left( \Delta m_{21}^2/\Delta m_{32}^2 \right)^{1/4} \right] $ in order to express the Yukawa couplings $y_\Omega^i$ and $y_\Delta^i$ with $i=1,2,3$ in term of the quantities $\hat{y}_\Omega$ and $\hat{y}_\Delta$ where 
\begin{equation}
    \hat{y}_\Omega \, \equiv \, \dfrac{\sqrt{\Delta m_{21}^2}}{2 \hat{y}_\Delta \sin^2(\phi_N)  m_f F_{\rm loop}  (m_{S^0_{1,2}},m_{S^\pm_{1,2}},m_f)}\, ,
\end{equation}
by the relations
\begin{align}
    \begin{pmatrix} y_\Delta^1 \\  y_\Delta^2 \\ y_\Delta^3 \end{pmatrix} & = \hat{y}_\Delta \, R_{\nu} \, \begin{pmatrix} 0 \\  \sin\phi_N \\ \cos\phi_N \end{pmatrix} \, , \\
    \begin{pmatrix} y_\Omega^1 \\  y_\Omega^2 \\ y_\Omega^3 \end{pmatrix} & = \hat{y}_\Omega \, R_{\nu} \, \begin{pmatrix} 0 \\  - \sin\phi_N \\ \cos\phi_N \end{pmatrix} \, ,
\end{align}
where the rotation matrix $R_{\nu}$ is written in terms of the observed neutrino mixing angles:
\begin{align}
      R_{\nu}&= \begin{pmatrix} 1 & 0 & 0 \\  0 & \cos\theta_{23} & \sin\theta_{23} \\ 0 & -\sin\theta_{23} & \cos\theta_{23} \end{pmatrix} \begin{pmatrix} \cos\theta_{13} & 0 & \sin\theta_{13} \\  0 & 1 & 0 \\ -\sin\theta_{13} & 0 & \cos\theta_{13} \end{pmatrix} \begin{pmatrix} \cos\theta_{12} & \sin\theta_{12} & 0 \\  -\sin\theta_{12} & \cos\theta_{12} & 0 \\ 0 & 0 & 1 \end{pmatrix} \, .
\end{align}

With this parametrization, neutrino masses can be expressed as
\begin{align}
    m_{\nu_1}\, = \, & 0 \, , \nonumber \\
    m_{\nu_2}\, = \, & -2 \hat{y}_\Delta \hat{y}_\Omega \sin^2(\phi_N) m_f F_{\rm loop}  (m_{S^0_{1,2}},m_{S^\pm_{1,2}},m_f)  \, , \\
    m_{\nu_3}\, = \, & -2 \hat{y}_\Delta \hat{y}_\Omega \cos^2(\phi_N) m_f F_{\rm loop}  (m_{S^0_{1,2}},m_{S^\pm_{1,2}},m_f) \nonumber \, .
\end{align}
The loop function $F_{\rm loop}$ can be expressed as
\begin{align}
    F_{\rm loop}&(m_{S^0_{1,2}},m_{S^\pm_{1,2}},m_f) = \nonumber \\ & -s_{\kappa} \sqrt{1 - \psi_0^2} \left(\frac{m^2_{S^0_1}}{m_f^2 - m^2_{S^0_1}} \log{\left(\frac{m^2_{S_1^0}}{m_f^2} \right)} - \frac{m^2_{S^0_2}}{m_f^2 - m^2_{S^0_2}} \log{\left(\frac{m^2_{S_2^0}}{m_f^2} \right)}\right) \nonumber \\
    & + s_{\kappa} \sqrt{1 - \psi_1^2} \left(\frac{m^2_{S^\pm_1}}{m_f^2 - m^2_{S^\pm_1}} \log{\left(\frac{m^2_{S_1^\pm}}{m_f^2} \right)} - \frac{m^2_{S^\pm_2}}{m_f^2 - m^2_{S^\pm_2}} \log{\left(\frac{m^2_{S_2^\pm}}{m_f^2} \right)}\right)\, .
\end{align}
Notice here that beside the scalar and vector-like state masses $m_f$, $\hat{y}_\Delta$ is the only free parameter of the neutrino sector.\par \medskip

To summarize, the set of independent variables used explicitly for the scan is 
\begin{equation}
    \{\kappa, s_\kappa, \lambda_{\Omega}, \lambda_{\Delta}, \lambda_{\Delta}^\prime,  \lambda_{H \Omega}, \lambda_{H \Delta}^\prime, \lambda_{ \Delta \Omega}, \phi_0, \phi_1,\hat{y}_\Delta, m_{S_1^0}, m_f\}\,.
\end{equation}

%%%%%%%%%%%%%%%%%%%%%%%%%%%%%%%%%%%%%%%%%%%%%%%%%%%%%%%%%%%%%%%%%%%%%%%%%%%%%
\section{Vector-like fermions: mass splitting}
\label{sec:fermion_mass_splitting}
The neutral and charged fermions belonging to the doublet $f$ are degenerated with a mass  $m_f$ at tree-level. However a small mass splitting is generated by radiative corrections~\cite{Sher:1995tc}. The mass splitting corresponds to
\begin{equation}
    \Delta m_f = m_{f^\pm} - m_{f^0} = m_f \, \frac{\alpha_{\rm EM}}{2 \pi} \Pi\left(\frac{m_{Z}}{m_f}\right) \, ,
\end{equation}
where $\alpha_{\rm EM}$ is the fine structure constant, $m_Z$ is the $Z-$boson mass, and $m_f$ is the tree-level mass of vector-like fermions. The $\Pi$ function can be expressed as
\begin{equation}
    \Pi\left(r\right) \, \equiv \, \int_{0}^1 \diff x (1-x) \left( \log{\left( x^2 + r^2 (1-x)\right)} - \log{\left( x^2\right)}\right) \, .
\end{equation}
This corresponds to the functional form produced by the radiative correction of the photon diagram that only affects the charged component of the vector-like fermion. This function can be reduced to the following analytical expression
\begin{equation}
    \Pi\left(r\right) = \left\{ 
    \begin{array}{ll}
         \frac{1}{4} r \left( 2 r (r^2 \log{r} - 1) + 2 \sqrt{|r^2 - 4|} (r^2 +2) \arctan{\left(\dfrac{\sqrt{|r^2 - 4|}}{r}\right)}  \right) \, , & {\rm for}\, \, 0<r<2 \\ \\
         \frac{1}{4} r \left( 2 r (r^2 \log{r} - 1) + \sqrt{r^2 - 4} (r^2 +2) \log{\left( \dfrac{r - \sqrt{r^2 - 4}}{r + \sqrt{r^2 - 4}} \right)}\right) \, , & {\rm for} \,\, 2\leq r 
    \end{array}
    \right. \, .
\end{equation}
%%%%%%%%%%%%%%%%%%%%%%%%%%%%%%

\section{Lepton flavour violation: computation of $\mu \rightarrow e \gamma$}
In our model, contributions to $\mu \rightarrow e \gamma$ at the loop level are given by diagrams where a photon is emitted by a charged particle running in the loop as shown in Fig.~\ref{fig:muegamma}. There are 8 diagrams where the charged particles in the loop can be $S_{1,2}^\pm$, $S^{\pm \pm}$, $f^+$, $(f^+)^c$. The functional dependence of the diagrams for which the photon is emitted by a scalar or fermionic states are encoded by the functions $F$ and $G$ respectively and were derived using the \texttt{PackageX} code \cite{Patel:2015tea,Patel:2016fam}. The coefficient of the operator responsible for $\mu \rightarrow e \gamma$, appearing in Eq. (\ref{eq:expressionBRmutoegamma}), can be expressed in terms of these functions as
\begin{align}
     \mu_{\mu e} \,=\,\frac{e \, m_\mu   }{64 \pi ^2} & \left[
     y_{\Delta}^2 y_{\Delta}^1 \left( \dfrac{\cos^2 \theta_1}{m_{S_{2}^\pm}^2} F \left( \dfrac{m_{f}^2}{m_{S^\pm_2}^2} \right) + \dfrac{\sin^2 \theta_1}{m_{S_{1}^\pm}^2} F \left( \dfrac{m_{f}^2}{m_{S^\pm_1}^2} \right) +\dfrac{4}{m_{S^{\pm \pm}}^2} F \left( \dfrac{m_{f}^2}{m_{S^{\pm \pm}}^2} \right) \right. \right. \nonumber \\ 
     & \left. \left.  +\dfrac{2}{m_{S^{\pm \pm}}^2} G \left( \dfrac{m_{f}^2}{m_{S^{\pm \pm}}^2} \right)
     \right)  +  y_{\Omega}^2 y_{\Omega}^1 \left(   \dfrac{\cos^2 \theta_0}{m_{S_{1}^0}^2} G \left( \dfrac{m_{f}^2}{m_{S^0_1}^2} \right) + \dfrac{\sin^2 \theta_0}{m_{S_{2}^0}^2} G \left( \dfrac{m_{f}^2}{m_{S^0_2}^2} \right) \right. \right.\nonumber \\
     \,& \left.  \left. +\dfrac{2\cos^2 \theta_1}{m_{S_{1}^\pm}^2} F \left( \dfrac{m_{f}^2}{m_{S^\pm_1}^2} \right) + \dfrac{2\sin^2 \theta_1}{m_{S_{2}^\pm}^2} F \left( \dfrac{m_{f}^2}{m_{S^\pm_2}^2} \right)
     \right)
     \right]\,,
     \label{eq:mutoegamma_mufactor}
\end{align}
where the function $F$ is defined as
\begin{equation}
    F(x)\,\equiv \,\frac{2 x^3+3 x^2-6 x^2 \log (x)-6 x+1}{6 (x-1)^4}\,,
\end{equation}
and the function $G$ is related to the function $F$ by
\begin{equation}
    G(x)\,\equiv \, \frac{x^3-6 x^2+3 x+6 x \log (x)+2}{6 (x-1)^4}\,=\,\dfrac{F(x^{-1})}{x}\,.
\end{equation}

%%%%%%%%%%%%%%%%%%%%%%%%%%%%
\section{Additional plots}
\label{sec:additional_plots}
\begin{figure}[!ht]
\centering
    \includegraphics[width=0.48\textwidth]{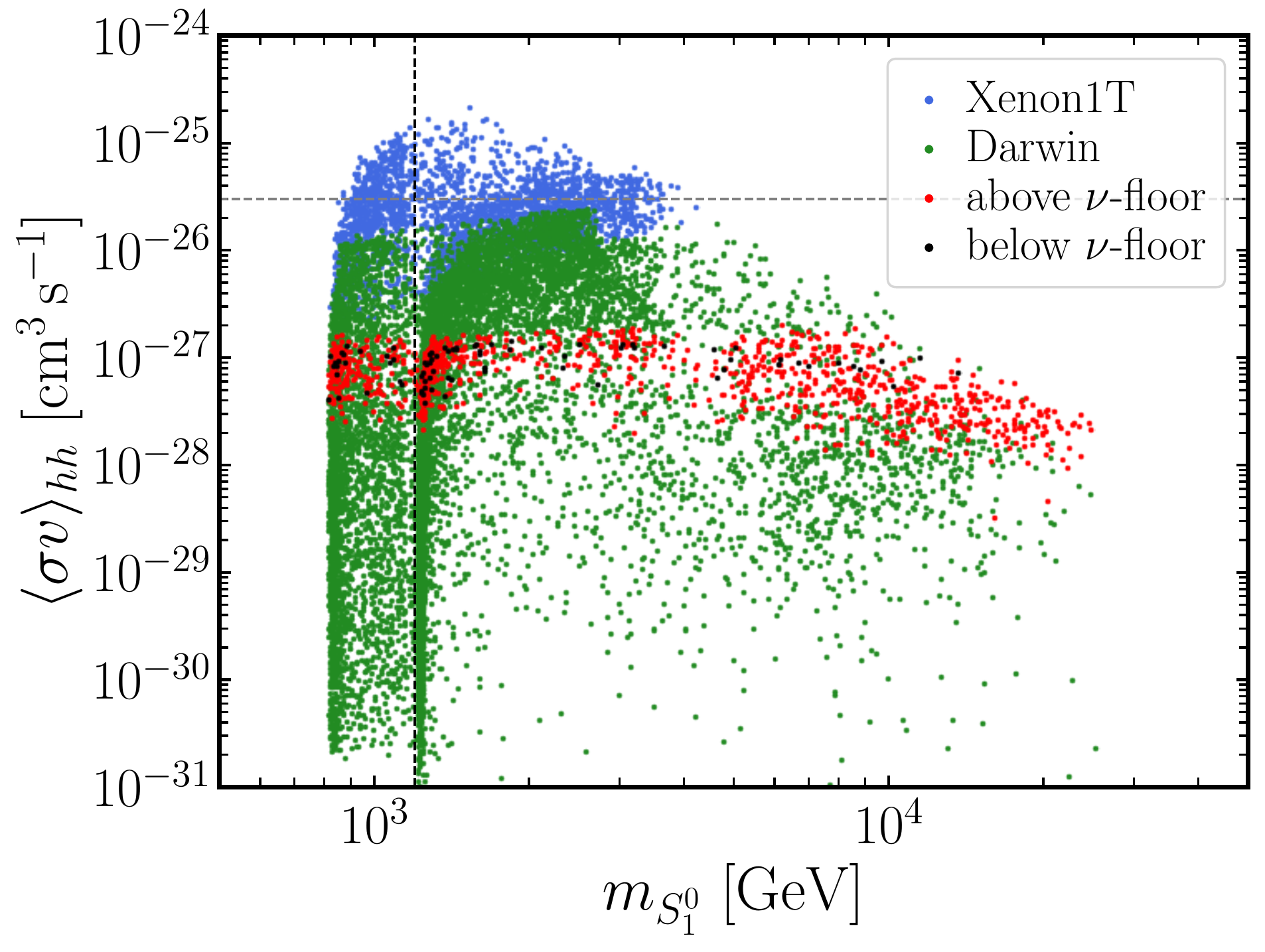}\hfill
        \includegraphics[width=0.48\textwidth]{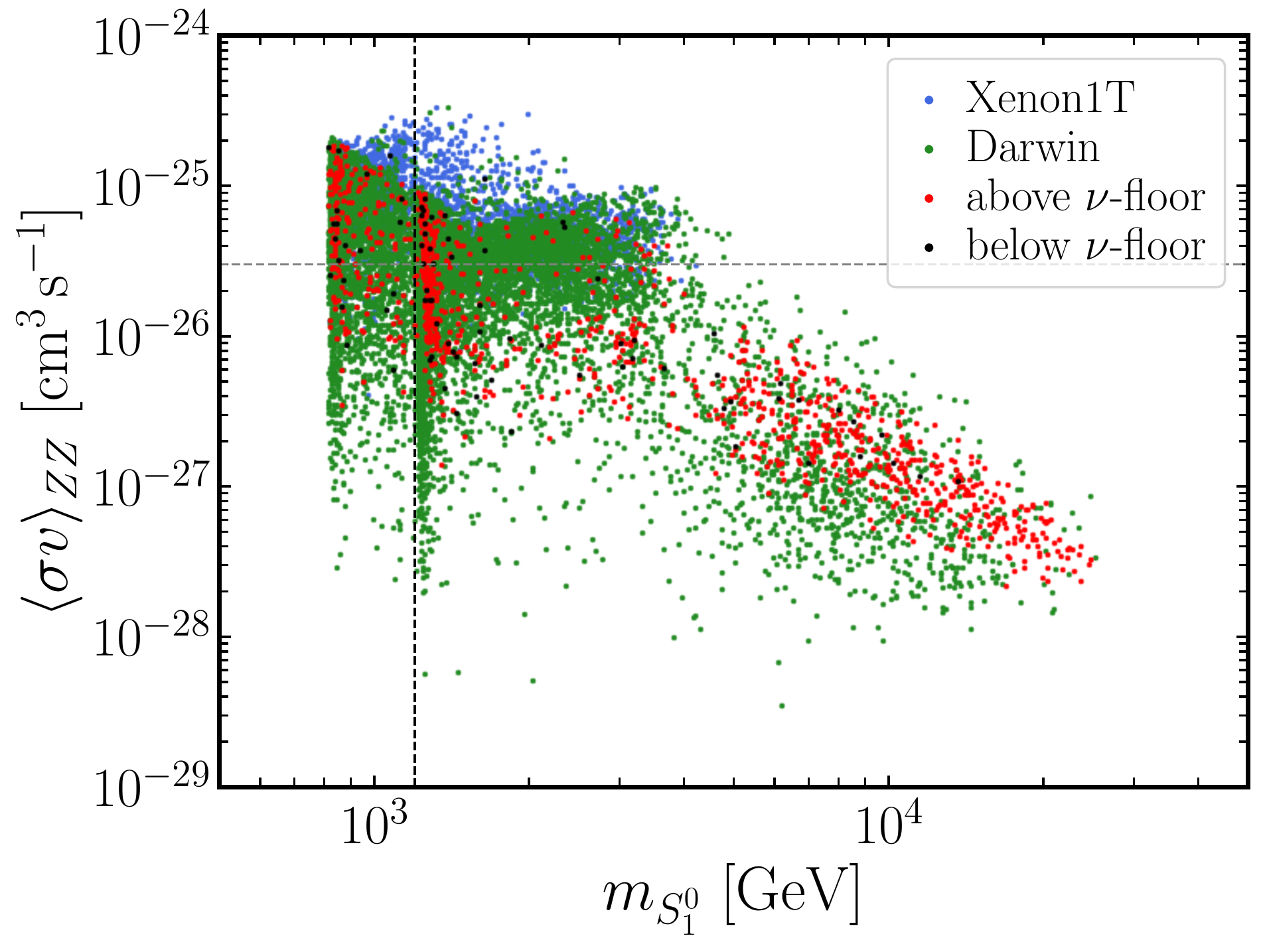}
    \caption{Scan in the parameter space: dark matter velocity averaged annihilation cross section into $hh$ (left) and $ZZ$ (right).  The color code is the same as in Fig.~\ref{fig:DD}, as described in Sec.~\ref{sec:DD}.}
    \label{fig:IDbosons}
\end{figure}

\begin{figure}[!htb]
\centering
    \includegraphics[width=0.47\textwidth]{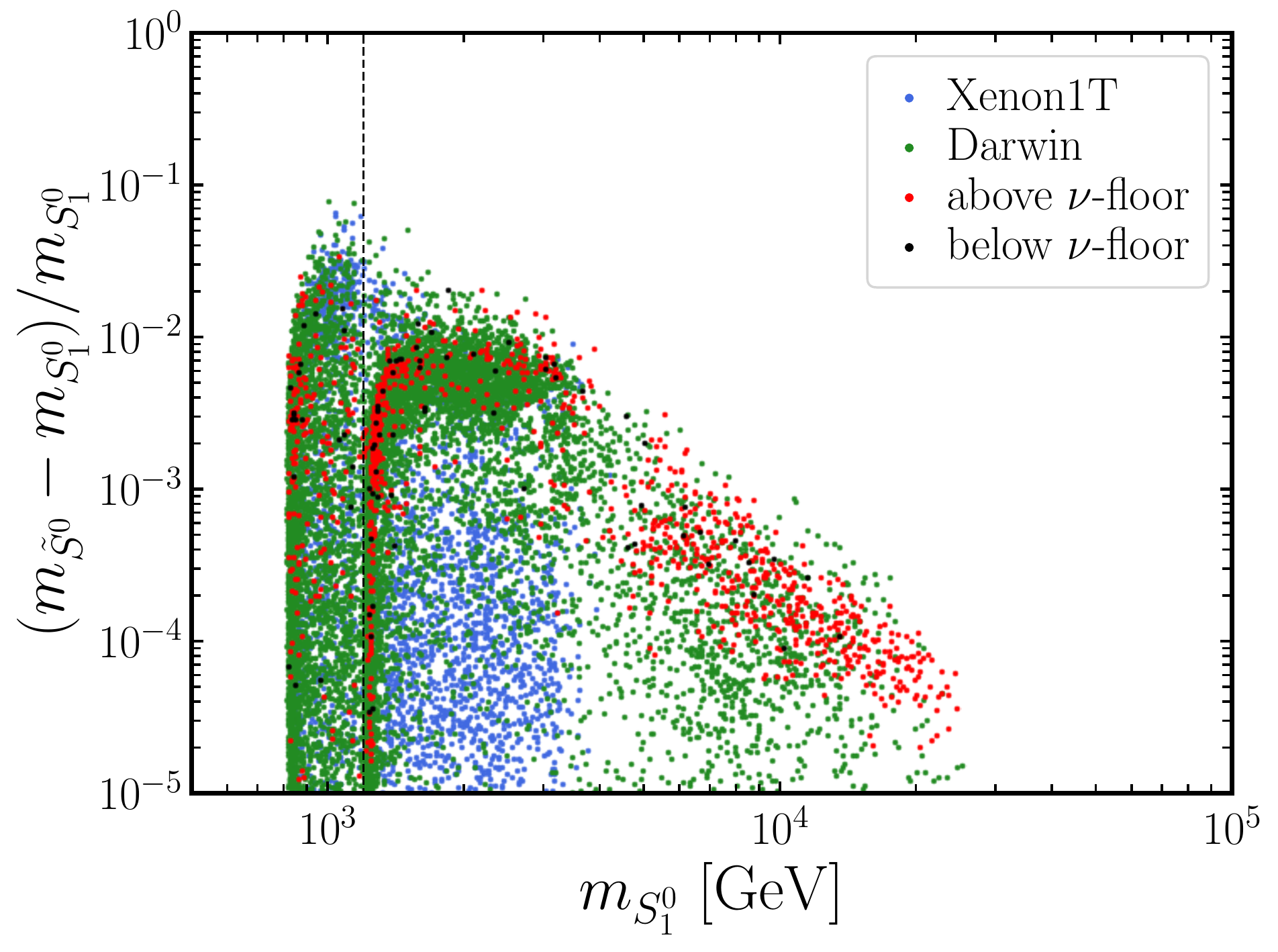}
    \hfill
    \includegraphics[width=0.47\textwidth]{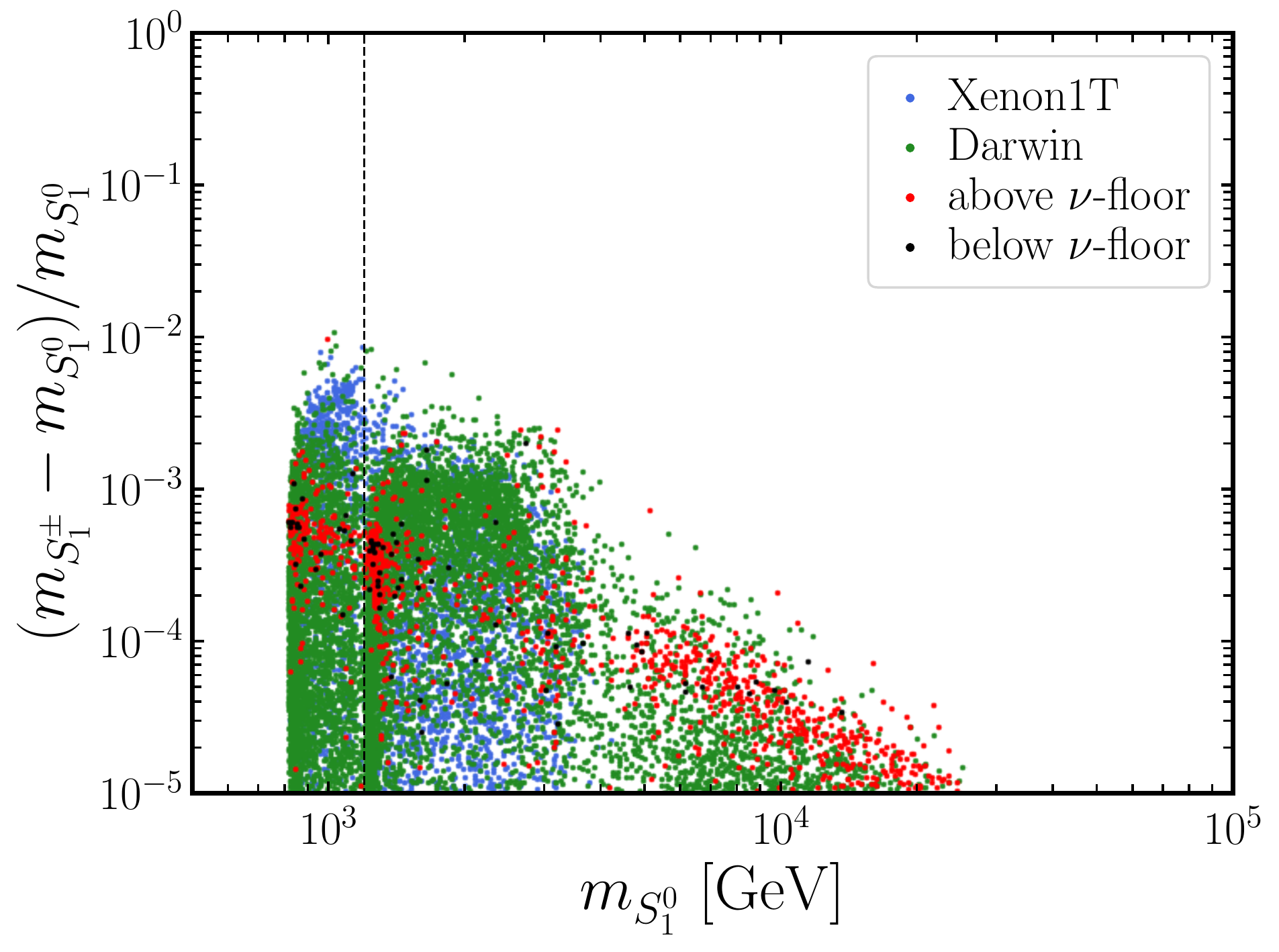}
    \caption{Scan in the parameter space: relative mass splitting between the neutral pseudoscalar (left), the lightest charged scalar (right) and the dark matter candidate. The color code is the same as in Fig.~\ref{fig:DD}, as described in Sec.~\ref{sec:DD}.}
    \label{fig:mass_splitting}
\end{figure}

\begin{figure}[!htb]
\centering
    \includegraphics[width=0.47\textwidth]{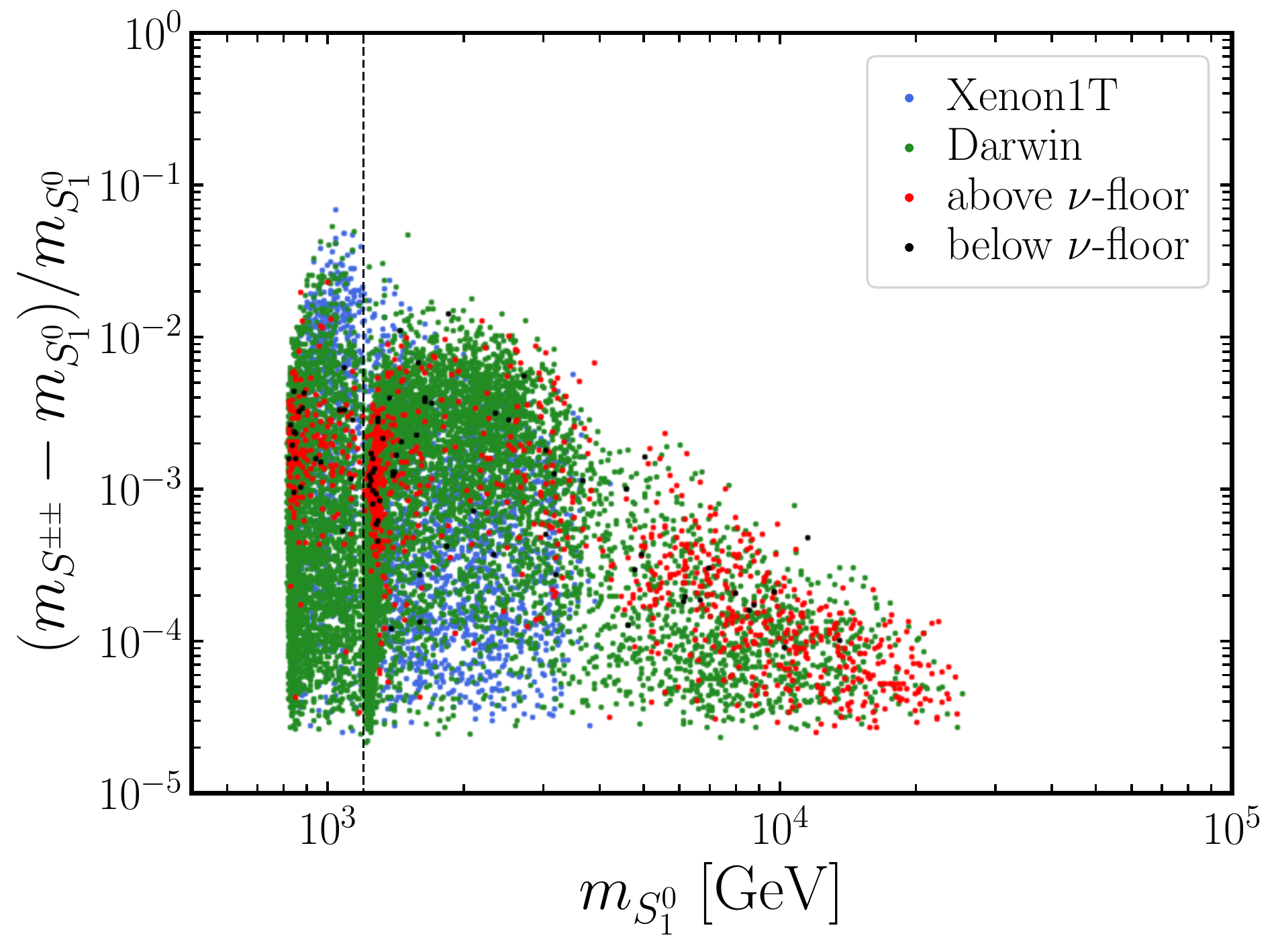}
    \hfill
    \includegraphics[width=0.47\textwidth]{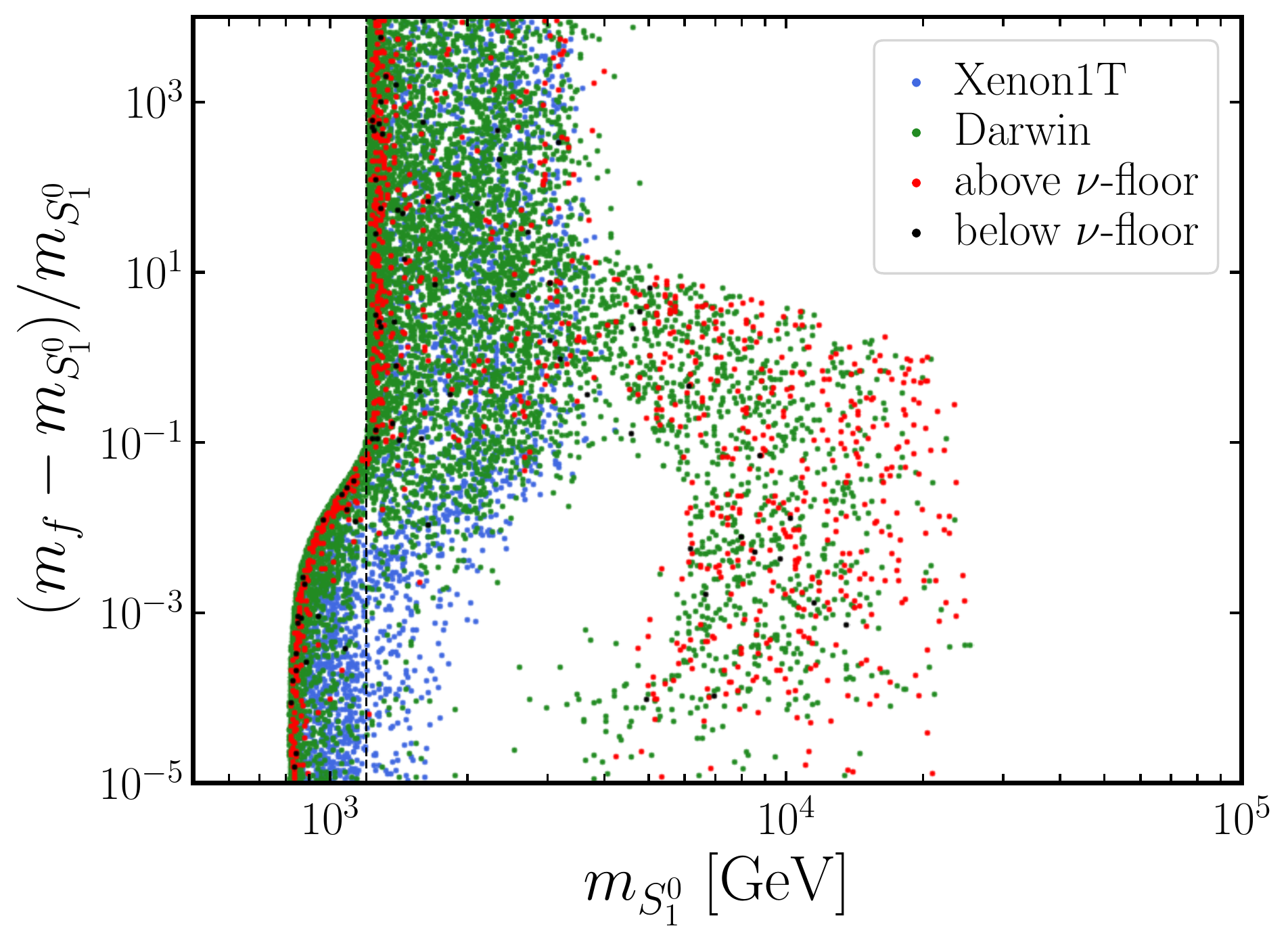}
    \caption{Scan in the parameter space: relative mass splitting between the charge-2 scalar (left), the fermionic state (right) and the dark matter candidate. The color code is the same as in Fig.~\ref{fig:DD}, as described in Sec.~\ref{sec:DD}.}
    \label{fig:mass_splitting2}
\end{figure}

\begin{figure}[!htb]
\centering
    \includegraphics[width=0.48\textwidth]{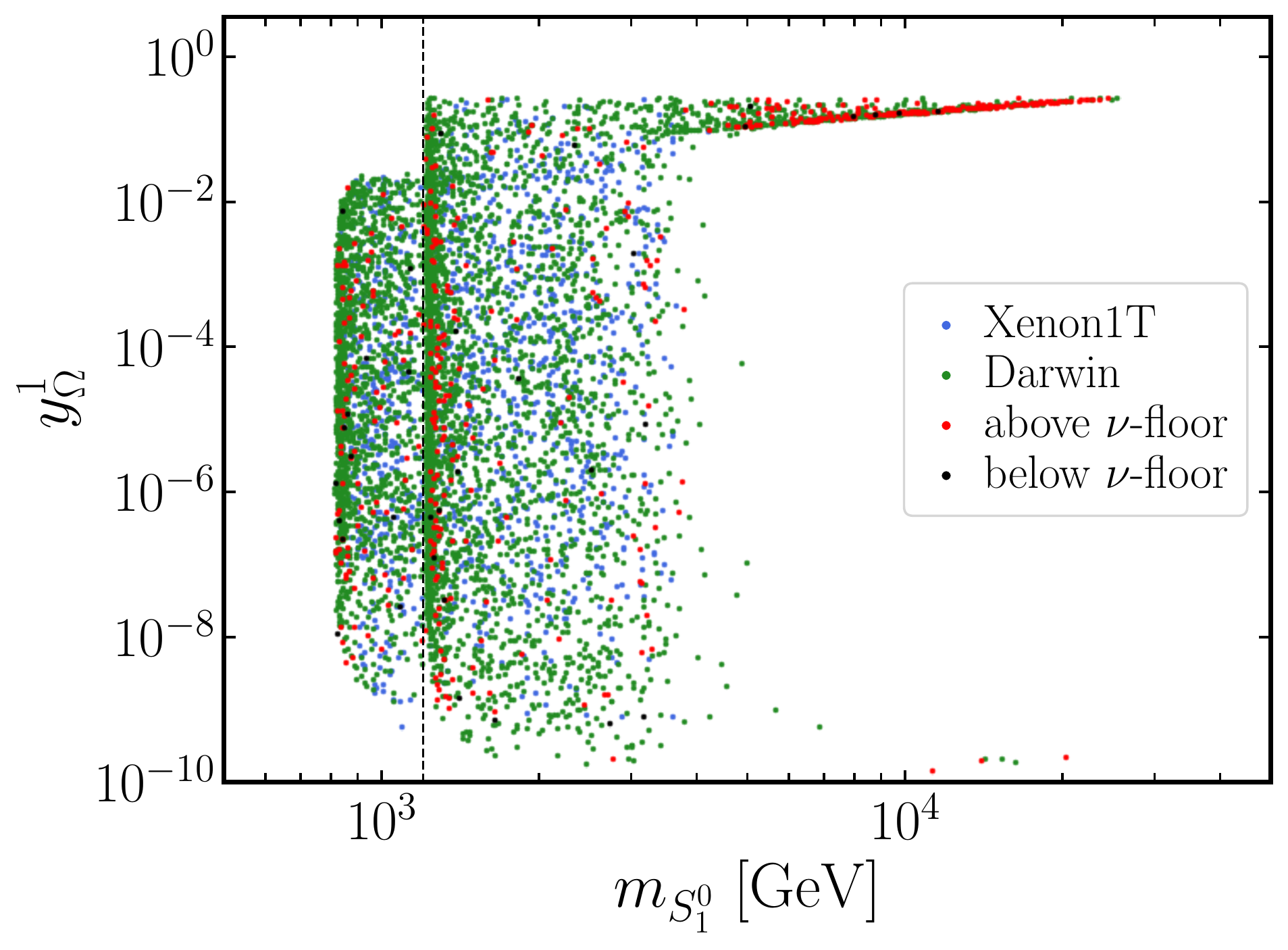}
    \includegraphics[width=0.48\textwidth]{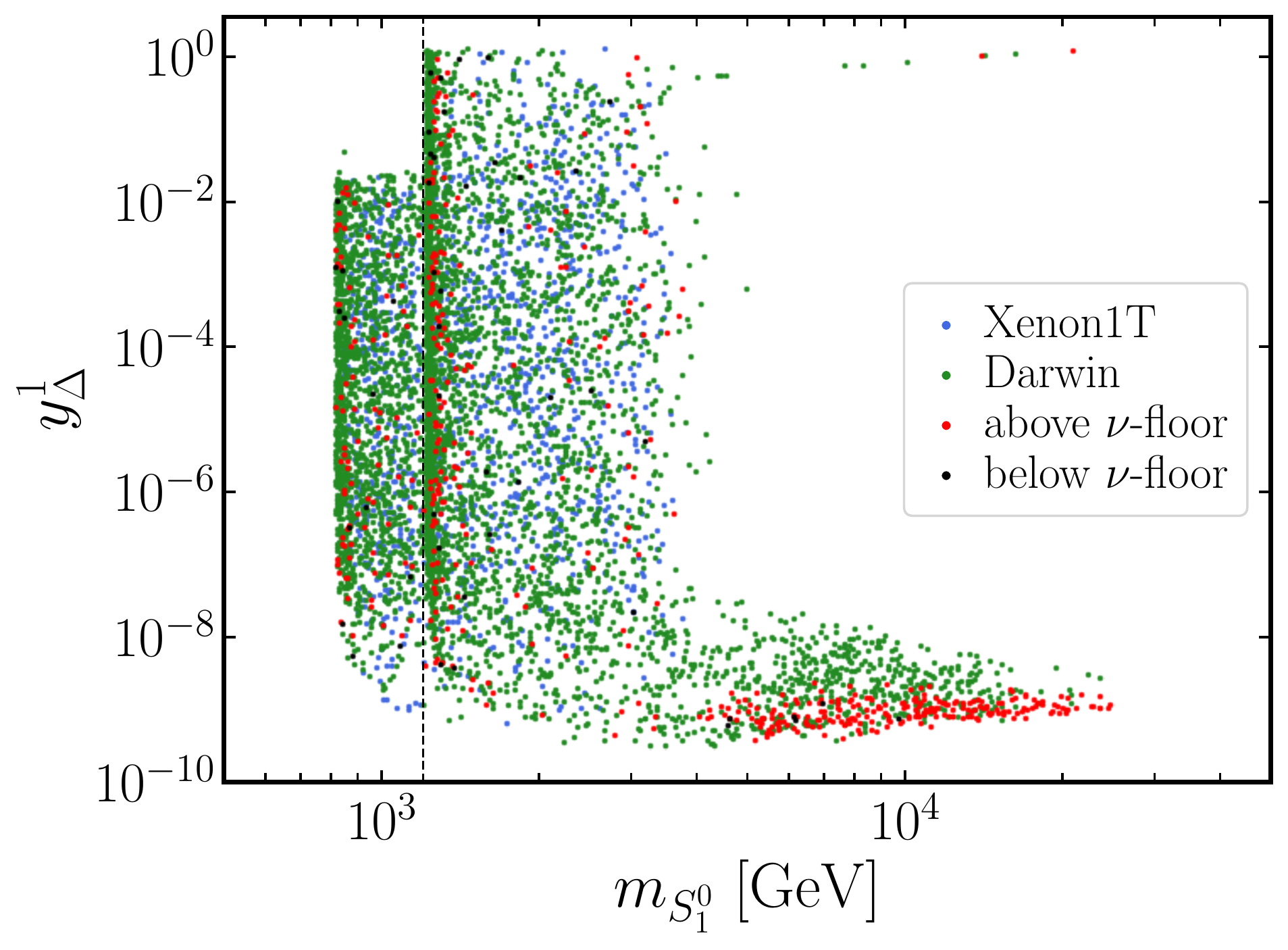}
    \caption{Scan in the parameter space: Yukawa couplings as a function of the DM mass. The color code is the same as in Fig.~\ref{fig:DD}, as described in Sec.~\ref{sec:DD}.}
    \label{fig:yukawacouplings}
\end{figure}

In this part of the appendix, we provide useful additional plots corresponding to our numerical scan in the parameter space as described in Sec.~\ref{sec:numerical_analysis}, to illustrate the interplay between the various couplings and observable quantities of this model. In Fig.~\ref{fig:IDbosons} we depicted the perturbative velocity averaged annihilation cross section into a pair of Higgs or $Z$-bosons of the SM. In Fig.~\ref{fig:mass_splitting} and Fig.~\ref{fig:mass_splitting2}, we represented the mass splitting between various scalar or fermionic mass eigenstates and the DM mass. In Fig.~\ref{fig:yukawacouplings}, we show predicted values of the Yukawa couplings $y_{\Omega}^1$ and $y_{\Delta}^1$.

\section{Diagrams for dark matter production}
\label{sec:diagrams}
\begin{figure}[hbt]
\centering
    \includegraphics[width=0.9\textwidth]{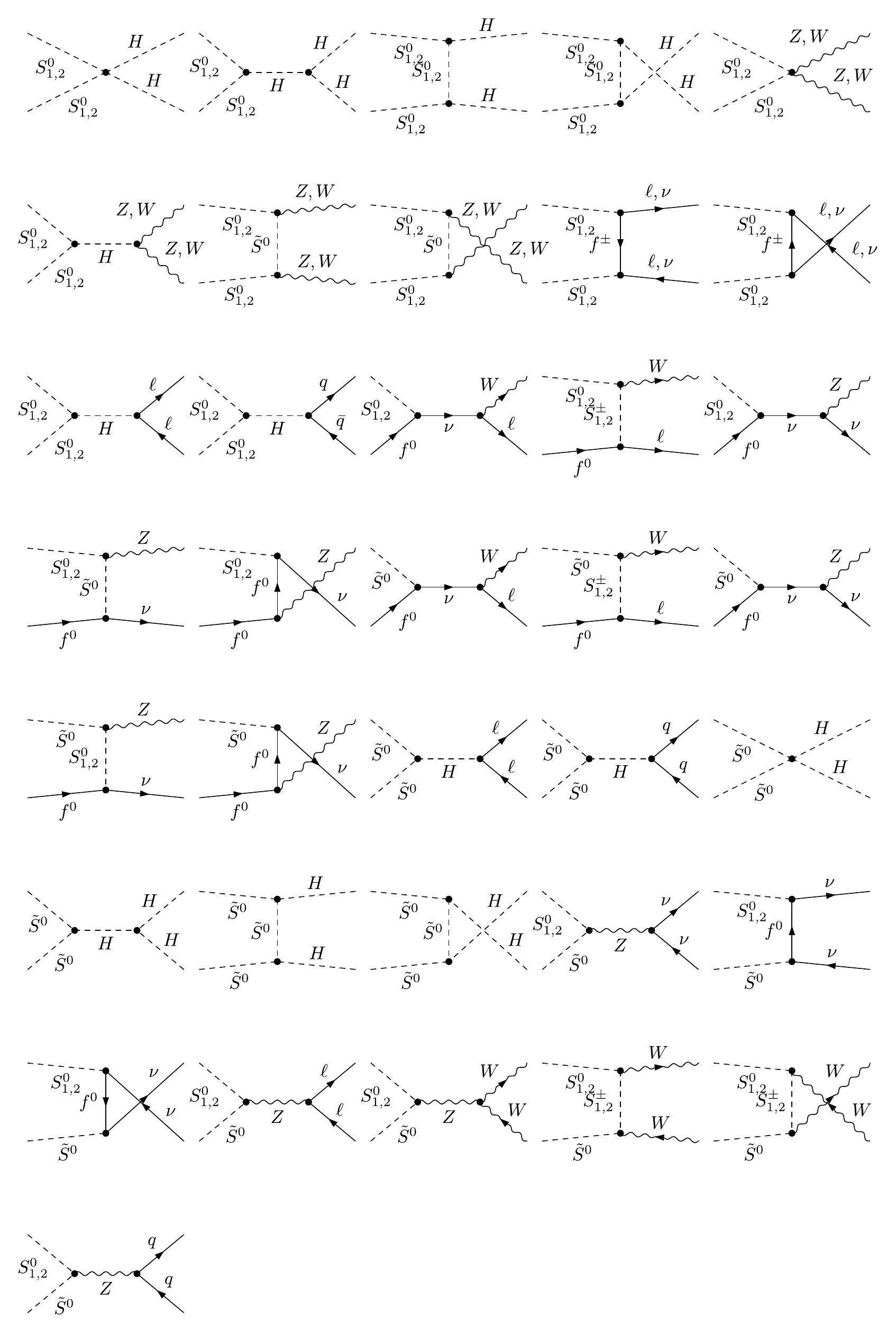}
    \caption{Diagrams relevant for the DM production involving neutral initial states ($Q=0$).}
    \label{fig:diagrams1}
\end{figure}

\begin{figure}[hbt]
\centering
    \includegraphics[width=0.9\textwidth]{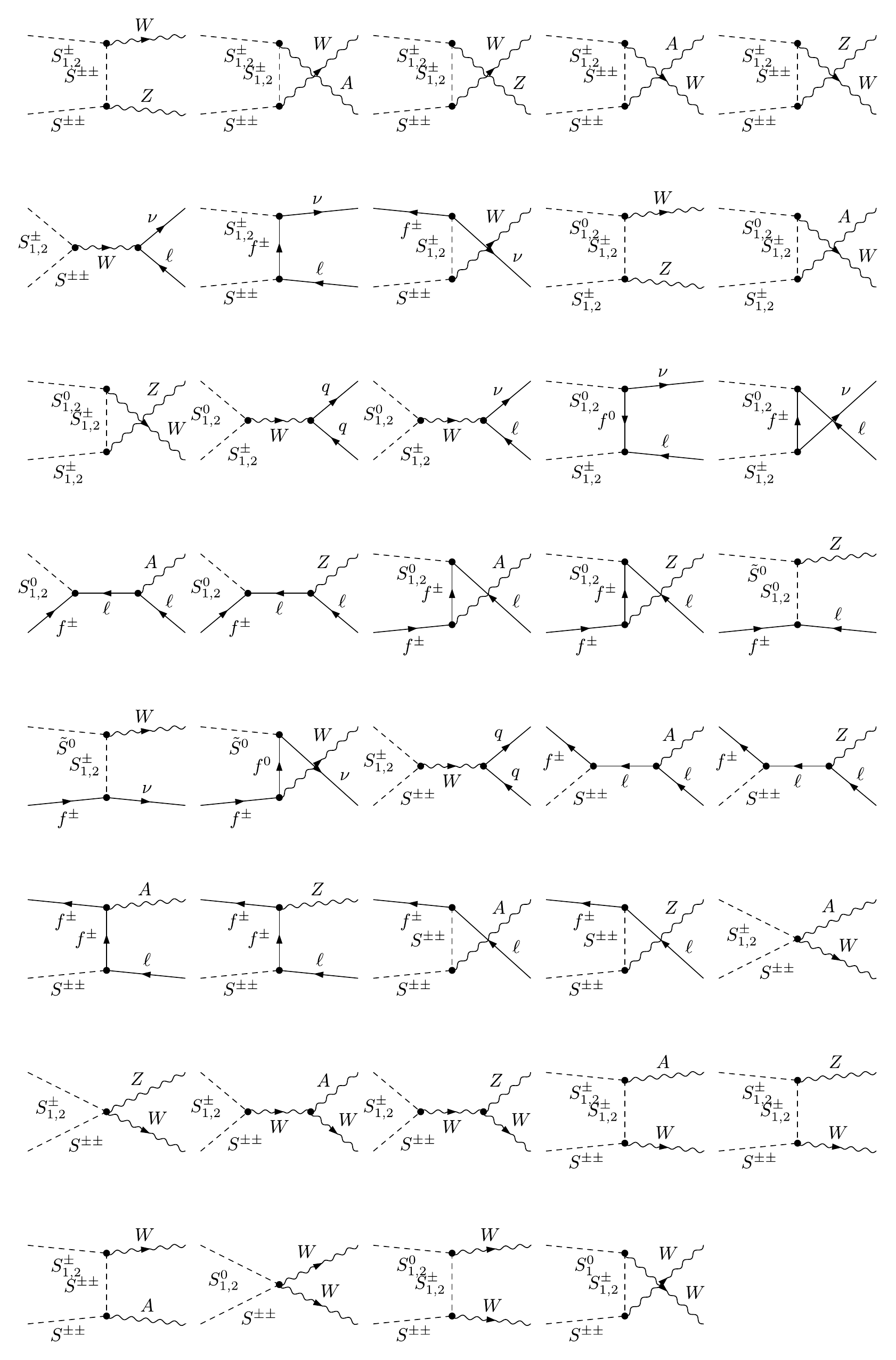}
    \caption{Diagrams relevant for the DM production involving charged initial states ($Q=1,2$).}
    \label{fig:diagrams2}
\end{figure}

In Fig.~\ref{fig:diagrams1} and Fig.~\ref{fig:diagrams2} we represented the numerous diagrams involved in annihilations and co-annihilations of $\mathbb{Z}_2$-odd states, relevant for the DM density production.

%%%%%%%%%%%%%%%%%%%%%%%%%%%%%%%%
%%%%%%%% REFERENCES
%%%%%%%%%%%%%%%%%%%%%%%%%%%%%%%%
%%%%%%%% Journals %%%%%%%%%%%%%%
\def\apj{Astrophys.~J.}                       % Astrophysical Journal
\def\apjl{Astrophys.~J.~Lett.}                % Astrophysical Journal, Letters
\def\apjs{Astrophys.~J.~Suppl.~Ser.}          % Astrophysical Journal, Supplement
\def\aap{Astron.~\&~Astrophys.}               % Astronomy and Astrophysics
\def\aj{Astron.~J.}                           %
\def\araa{Ann.~Rev.~Astron.~Astrophys.}       %
\def\mnras{Mon.~Not.~R.~Astron.~Soc.}         %
\def\physrep{Phys.~Rept.}                     % Physics Reports
\def\jcap{J.~Cosmology~Astropart.~Phys.}      % Journal of Cosmology and Astroparticle Physics
\def\jhep{J.~High~Ener.~Phys.}                % Journal of High Energy Physics
\def\prl{Phys.~Rev.~Lett.}                    % Physical Review Letters
\def\prd{Phys.~Rev.~D}                        % Physical Review D
\def\nphysa{Nucl.~Phys.~A}                    % Nuclear Physics A
%\nocite{*}
\bibliographystyle{JHEP}
\bibliography{refs}

\end{document}